\documentclass[12pt]{article}%
\usepackage{amsmath}%
\usepackage{txfonts}%
\usepackage{amsfonts}%
\usepackage{amssymb}%
\usepackage{graphicx}
\usepackage{color}
\usepackage{setspace}
\usepackage{esint}
\usepackage[rightcaption]{sidecap}
\usepackage{wrapfig}
\usepackage{lipsum}
\usepackage{caption}
\usepackage{longtable}
\usepackage[T2A,T1]{fontenc}
\usepackage[utf8]{inputenc}
\usepackage[russian,english]{babel}

\numberwithin{equation}{section}

\newcommand{\be}{\begin{equation}}
\newcommand{\ee}{\end{equation}}

\usepackage{anysize}
\marginsize{2.5cm}{2.5cm}{1cm}{2cm}
\usepackage{mathptmx}
-1in\relax 

\begin{document}
\LARGE
\begin{center}
\textbf{Gas Core Reactors for Deep Space Propulsion}
\end{center}
\normalsize
\vskip1cm
\begin{center}
{\Large \textbf{Mikhail V. Shubov}}
\normalsize
\vskip1cm

University of MA Lowell,\quad
One University Ave,\quad
Lowell, MA 01854

E-mail: mvs5763@yahoo.com  \\
\end{center}

\begin{center}
  \textbf{Abstract}
\end{center}

The operation of a nuclear power station powering Deep Space transport is presented in the paper.  The power station consists of a Gas Core Reactor (GCR) with magnetohydrodynamic (MHD) generator.
A new model of gas core reactor which has advantages by comparison with known models is suggested.
It is shown that $200\ ton$ system would generate $200\ MW$ electricity.  The gas core reactor consists of a spherical cavity of $1\ m$ to $1.5\ m$ radius surrounded by a $50\ cm$ thick liquid beryllium reflector within molybdenum--92 container.  The fuel consists of $^{235}$UF$_4$, KF, and K.  It is demonstrated that such power station will work for gaseous fuel pressures of $50\ atm$ to $100\ atm$.  The Monte-Carlo simulations describing paths of thousands of neutrons are performed.
These simulations yield neutronic results, i.e. the fractions of neutrons causing fission,  being absorbed by moderator, and escaping the reactor.  From this information one can calculate minimum needed pressure inside the reactor, the reactor size, and moderator thickness.
The conductivity of superheated fuel depending on temperature and pressure is also calculated.  It is concluded that space propulsion is the best use of uranium resources.

\begin{center}
  \textbf{Introduction}
\end{center}
\hskip.5cm Deep Space transportation of large payloads would use electrically powered magnetoplasmadynamic thrusters to provide momentum for maneuvers.  These thrusters use lithium propellant.  The thrusters should provide a sustained acceleration of
$10^{-3}\ m/s^2$ for the space transport.  Although this value seems incredibly small compared to planetary gravity, it would certainly be sufficient for most maneuvers of interplanetary travel.  As we show in Subsection 1.3, the minimum electrical power requirement for the thrusters is $45\ W/kg \cdot M_{\text{tot}}$, where $M_{\text{tot}}$ is the total mass of the space transport.  The space transport includes an electrical power source, the thrusters, lithium propellant, payload, and the spaceship.  Thus, the space transport must contain an electrical energy source of high specific power.

Several technologies exist for generating electrical energy in outer space in general and deep space in particular.  They are photovoltaic arrays, solid core fission reactors with turbine generators, and gas core reactors (GCR) with magnetohydrodynamic (MHD) electrical generators \cite{Espace}.  In this work we describe a power system consisting of a GCR with a MHD electrical generator.  The power system will generate about $1,600\ MW$ thermal power and convert $200\ MW$ into electric power, which means that thermal efficiency will be 12.5\%.  The system with all accessories will have a mass of about $200\ tons$ and thus a specific electric power of about $1\ kW/kg$.  If we scale down such a system, the specific electric power will have to be much less.   All other power systems have specific power at least 5 times lower \cite[p.3]{Espace}.  As we will show in Section 6, the cost of such system will be \$ 4.2 Billion.

The Gas Core Reactor will consist of a spherical or cylindrical cavity surrounded by liquid beryllium reflector in molybdenum--92 container.  The reactor cavity radius will be from $1.0\ m$ to $1.5\ m$ and beryllium reflector thickness will be $50\ cm$.  The fuel consisting of $^{235}$UF$_4$, KF, and K in gaseous form will enter the reactor cavity at $50\ atm$ -- $100\ atm$ pressure.  It will be heated by $^{235}$U fission to 3,500$\ ^o K$.  Then the fuel gas will expand and generate electric power in a magnetohydrodynamic generator.  Finally, the fuel will be cooled and condensed in radiative heat rejection system.

The GCR-MHD system was originally suggested in \cite{MHD03,UF4-01}.  We introduce several modifications in the construction and work of the system.  First, we lower the working temperature of the gas from 4,000 $^oK$ to 3,500 $^oK$.  Since 4,000 $^oK$ exceeds the melting point of tungsten, reactor working at such temperature may be worn out very quickly.  Second, we suggest a different neutron moderator.  In \cite{MHD03,UF4-01} it is beryllium oxide.  We propose to use liquid beryllium within molybdenum tubing.  That would greatly simplify the cooling of moderator and slightly improve the neutron balance.  Unlike the previous works, we have performed calculations of neutron balance using extensive Monte Carlo simulations.  We have shown that tungsten can not be used as a GCR shell, while molybdenum can.

Now we are in a position to outline the content of the present work.
In Section 1 we present a brief review of possible means of propulsion for deep space transportation.  First we discuss the electric propulsion.  The basic engine for space propulsion is the magnetoplasmodynamic (MPD) thruster.  At high power, MPD thrusters have many advantages.  Namely, they have simple construction, about 50\% efficiency, and high specific thrust.  They use inexpensive propellants like lithium and sodium.  Another engine is called Variable Specific Impulse Magnetoplasma Rocket (VASIMR) \cite{Vasya}.  VASIMR is also efficient, but it is much more complicated and expensive then MPD.  It is useful only for light spacecraft.  Second, we describe thermal nuclear propulsion (TNP).  The advantage of TNP is that it uses space-based resources such as water and requires low specific power.  The disadvantage of TNP is low specific impulse.  Third, we calculate the power requirements for electrical and thermal propulsion.  Generally, electric propulsion requires 45 $kW$ per 1 $N$ of thrust.  That translates to 230 -- 450 $kW$ thermal power per Newton of thrust.  TNP uses at least 100 times less energy, but it is unusable on most space missions, because it requires a lot of neutral propellant like water which is not available at most space destinations.

In Section 2 we give an overview of the concept of a gas core reactor (GCR).  GCR consists of a cavity filled with fissile gas and surrounded by neutron reflector.  As fission chain reaction takes place in the reactor cavity, the gas is heated to 3,500 $^oK$.  Then the hot gas expends and generates electrical energy via a magnetohydrodynamic generator.  The fuel is passed through the reactor cavity thousands of times.  The gas is a mixture of $^{235}$UF$_4$, KF, and K.  The neutron reflector is liquid beryllium within molybdenum tubing.  The reactor is presented in Figure 3.  Previous models of GCR considered gas at 4,000 $^oK$.  In our opinion, such heat will destroy the generator very quickly.

In Section 3 we present the calculations of neutron balance.  These calculations determine under which conditions the reactor can achieve criticality.  The varying conditions include different reactor cavity sizes, tubing materials, reflector wall thicknesses, and fuel pressure.  Each neutron starts its path from a $^{235}$U fission, undergoes a number of collisions, and finally ends its path either by being absorbed or by escaping the reactor.  Each neutron travels through the reactor's three regions -- gaseous fuel, refractory shell, and reflector.  We present mathematical formulae describing neutron paths.  We derive neutron velocity after elastic and inelastic scattering.  Finally, we calculate the neutron balance and reactor criticality.  We calculate the impact of neutron absorbtion by fission products and neutron multiplication in beryllium.  The two aforementioned factors almost cancel each other.

In Section 4 we present the calculation results.  First, we present the fraction of neutrons reflected by a liquid beryllium reflector.  In some cases, over 90\% of neutrons are reflected.  Second, we show that a reactor with tungsten refractory shell can never achieve criticality.  Third, we show that a reactor with molybdenum-92 refractory shell can easily achieve criticality.  The lower melting point of molybdenum would present an engineering problem.

In Section 5 we calculate plasma density and conductivity for different temperatures and pressures.  First, we express plasma conductivity in terms of electron collision cross-sections.  Given that these cross-sections for $UF_4$, $UF_5$, and $UF_6$ are still untabulated, some experimental work remains to be done.  We show that conductivity sufficient for a magnetohydrodynamic generator can be achieved by expanding plasma.  In some cases, the expanding gases may have to be mechanically heated in order for conductivity to stay high.  Mechanical heating is accomplished by placing tungsten obstacles in the path of the high velocity gas stream.  Second, we calculate the generator efficiency.  It turns out to be 12.5\%.  Third, we calculate the specific power of the radiator heat rejection system.

In Section 6 we discuss economic considerations.  First we list the state of the art deep space launch costs.  Falcon by SpaceX is the most cost-efficient launch system.  It can launch payload into Low Earth Orbit (LEO) for \$5,400 per $kg$.  Second we estimate the cost of the reactor itself.  The most expensive material used in the reactor is the fuel -- \$89,000 per kilogram of $^{\text{235}}$U.  We estimate the cost of the whole reactor at about \$4.2 billion.  One of the factors lowering reactor cost is the much lower safety requirement compared to reactors on Earth.  Third we calculate the fuel efficiency of GCR.  The reactor will use 40\% of the fissile isotope $^{235}$U and discard 60\%.  This will have to be done due to accumulation of fission products and $^{236}$U in the uranium fuel.  As we show in Subsection 6.1, one gram of input $^{\text{235}}$U would produce $1,080\ KWh$ electrical energy, which is much less than $8,400\ KWh$ produced in modern pressurized water reactors \cite[p. 111]{allnuclear}.  Nevertheless, given the cost of delivering materials into deep space this is an outstanding performance.

Finally, we argue that the best use of uranium resources is for space propulsion.  The energy density of $^{235}$U fuel is 68 $GJ/g$ making it perfect for space applications.  For terrestrial applications, coal reserves have 21 times as much energy as uranium reserves.

\section{Deep Space transportation}

\hskip.8cm Using conventional chemical rocket fuel delivered from Earth for Deep Space colonization and exploration is cost prohibitive.  The cost of fuel and fuel tanks in Deep Space is at least \$20,000 per $kg$ due to the cost for bringing these materials from Earth \cite{Raport01}.  We explore two options for Deep Space propulsion -- electric propulsion and thermal propulsion.  Electric propulsion uses fuel brought from Earth and produces very high exhaust velocities.  Thermal propulsion uses materials obtained in space -- such as water (obtained, e.g., from ice on asteroids) and has moderate exhaust velocities.

\subsection{Electric propulsion}

\hskip.8cm The energy and fuel requirements for space missions or maneuvers are given in terms of
$\triangle v$ which is the change in velocity of the spacecraft needed to accomplish a given mission or maneuver \cite[pp.9--11]{spaceprop}. To evaluate the amount of fuel needed for a maneuver with $\triangle v$ we use the Tsiolkovsky's rocket equation \cite[p.9]{spaceprop}:
\be
\label{1.01}
\frac{m_0}{m_1}=\exp \left( \frac{\triangle v}{v_0} \right),
\ee
where $m_0$ is the mass of the rocket with fuel, $m_1$ is the mass of the vehicle after the fuel has been consumed, and $v_0$ is the exhaust velocity of the rocket engine.  The highest value of $v_0$ for a storable liquid rocket propellant, hydrazine and liquid fluorine, is $4.22\ km/s$ \cite{LRP}.  An obvious way to deliver greater mass over space routes demanding large
$\triangle v$ is to use engines with much greater exhaust velocity.  For low power applications \textbf{ion thrusters} and \textbf{Hall effect thrusters} \cite[p.5-6]{ion1} can be used.  For applications using hundreds of thousands of kilowatts or greater power, \textbf{magnetoplasmadynamic thruster} and Variable Specific Impulse Magnetoplasma Rocket (\textbf{VASIMR}) can be used.

Magnetoplasmadynamic thrusters pass current through a propellent and use Lorentz force to expel it at 15 to 60 $km/s$ \cite{MPT1}.  These thrusters generally have high power of 100 -- 500 $KW$.  The construction of these thrusters is quite simple \cite[p.15]{MPT2} and schematically shown on Fig. 1 below
  \begin{center}
  \includegraphics[width=5cm]{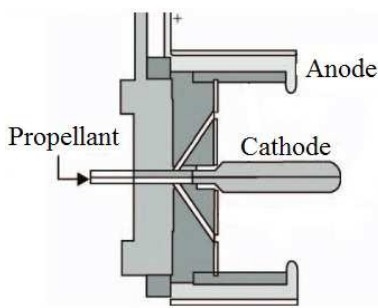}
  \vskip0cm
  \captionof{figure}{Magnetoplasmadynamic thruster}
  \end{center}
This thruster has not been used in space yet because of requirement of hundreds of kilowatts of electric power.  Sometimes the simplest ideas are the hardest to implement.  Variable specific impulse magnetoplasma rocket (VASIMR) ``is an electromagnetic thruster for spacecraft propulsion. It uses radio waves to ionize and heat a propellant. Then a magnetic field accelerates the resulting plasma to generate thrust (plasma propulsion engine)" \cite{Vasya}.

\hskip.5cm For an Ion thruster with power $P$, exhaust velocity $v_0$, and efficiency $\eta$, the following formula for the thrust force holds:
\be
\label{1.02}
F=\frac{2 P \eta}{v_0}.
\ee
The characteristics of several ion and Hall effect thrusters which have been used in space are given in \cite[p.57]{ion1}.  Those for magnetoplasmadynamic (MPD) thrusters are given in \cite[p.19]{MPT2}.  Those for VASIMR thrusters are given in \cite[p.7]{Vasya1}.
By measuring the size and estimating the volume of MPD thruster parts on diagram \cite[p.20]{MPT2}, we conclude that a $250\ kW$ thruster has a mass not exceeding $500\ kg$. Hence, MPD thrusters have specific mass of $2\ kg/kW$ or less.  In table below we have collected the data for electrical propulsion systems.
\begin{center}
\begin{tabular}{|c|c|c|c|c|c|c|}
  \hline
  Engine & Engine  & Propellant  & Power    & Thruster  &$v_0$ $km/s$   & Efficiency \\
  type   & name    &             & $kW$     & mass (kg)     &            &            \\
  \hline
  Ion    & T6  \cite[p.57]{ion1}    & xenon       & 6.8     &            &46            & .68  \\
  Ion    & NEXT \cite[p.2]{Next}    & xenon       & 6.9     &  5.0       &41            & .7  \\
  Ion    & HiPEP \cite[p.57]{ion1}  & xenon       & 30      &            &87            & .8  \\
  \hline
  Hall   & SPT-140 \cite[p.57]{ion1}  & xenon       & 5       &            &49            & .55 \\
  \hline
  VASIMR & VF-150 \cite[p.7]{Vasya1}  & argon       & 150     &   1,000         & 29            & .61 \\
  VASIMR & VX-200 \cite[p.9]{Vasya2} & argon       & 200     &    620  & 30 -- 50 & .6  \\
  VASIMR & VF-400 \cite[p.7]{Vasya1}  & argon       & 400     &   1,000          & 49            & .73 \\
  \hline
  MPD    & Fakel  \cite[p.19]{MPT2}  & lithium     & 300 -- 500 &  1000       & 34 -- 44 & .4 -- .6 \\
  MPD    & Energiya \cite[p.19]{MPT2}  & lithium    & 500        &  1000       & 44            & .55 \\
  \hline
\end{tabular}
\captionof{table}{Electric thrusters}
\end{center}
Scarcity of propellant presents a serious problem for Ion and Hall thrusters listed above.  Xenon is one of the rarest elements on Earth.  VASIMR and MPD thrusters listed use argon and lithium which are readily available.
The efficiency of VX-200 engine for different propellants is :
  \begin{center}
  \includegraphics[width=15cm]{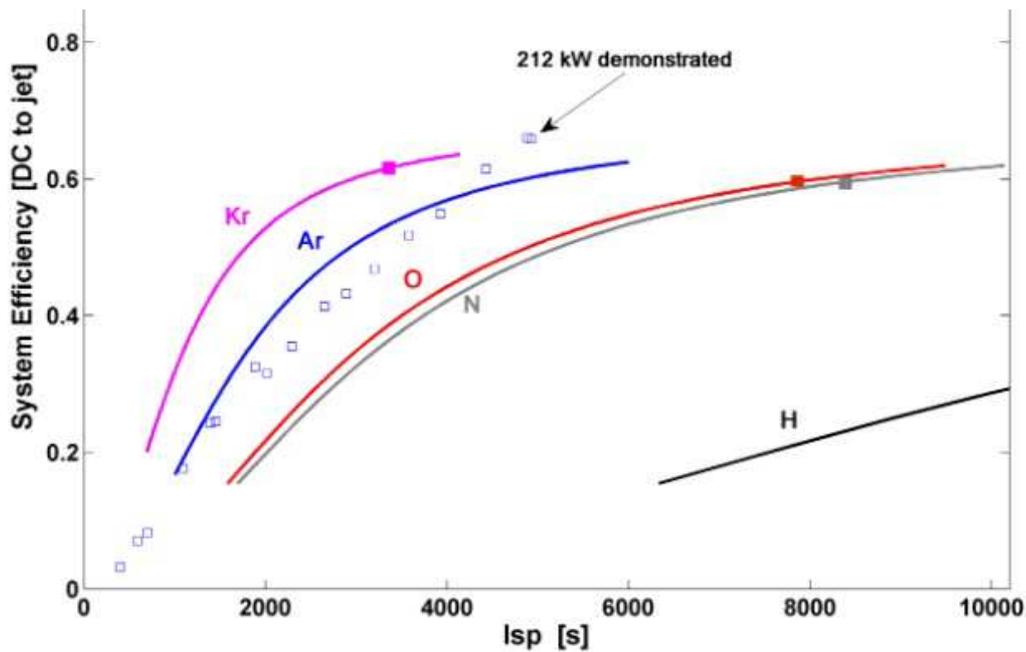}
  \vskip0cm
  \captionof{figure}{System efficiency of the VX-200 engine as a function of the specific impulse \cite[p.8]{Vasya3}.}
  \end{center}
Although krypton gives better performance than argon, it is too rare to be used on massive missions.  VASIMR engines are generally expensive e.g., a $200\ kW$ engine costs \$308 million and a $400\ kW$ engine costs \$371 million \cite[p.7]{Vasya1}, which means a cost of \$930 per $Watt$.  Unless VASIMR cost per $Watt$ can be decreased at least by a factor of 20 for multimegawatt thrusters, MPD thrusters present the only option available.

\subsection{Thermal nuclear propulsion}
\hskip.5cm A thermal nuclear rocket consists of a super hot reactor, which heats the propellant and a nozzle through which the heated propellant is expelled.  Propellant can consist of hydrogen, ammonia, hydrocarbons, or water.  These materials should be obtainable on asteroids.

In order to calculate performance for different propellant combinations at different temperatures, we have used the Rocket Propulsion Analysis program \cite{RPA}.  This program calculates exhaust velocity and chamber/exhaust composition for various propellant combinations.  The results are tabulated in Table 2 below.  The fourth column is the composition inside rocket chamber.  In all cases the area expansion ratio is 200.  The internal energy of propellant is the thermal energy it takes to heat the propellant to rocket chamber temperature and to accommodate all the phase and chemical changes.  The fifth column is thermal efficiency, i.e. the portion of thermal energy supplied to the propellant by heating transferred into kinetic energy.

\begin{center}
\begin{tabular}{|l|l|l|l|l|l|}
  \hline
  Propellant & Chamber     & Exhaust  & Chamber     & Internal & Thermal\\
             & temperature & velocity & composition & energy   & efficiency\\
             & $^oK$       & $m/s$    &             & $kJ/g$   & \\
  \hline
  hydrogen   & 1,800       & 7,030    & $H_2$       & 27.0    & 0.92\\
  \hline
  ammonia    & 1,900       & 3,530    & 17.7\% $H_2$,\ \ 82.3 \% $N_2$ & 8.3 & 0.75 \\
  \hline
  55\% water & 1,450       & 3,180    & 15\% $C H_4$,\ \ 44\% $CO$, & 7.3 & 0.69 \\
  45\% methane &           &          & 12\% $C O_2$,\ \ 6\% $H_2$,& &\\
               &           &          &   22\% $H_2 O$ & & \\
  \hline
  90\% methanol & 1,450    & 2,860    & 9\% $C H_4$,\ \ 59\% $CO$,& 4.45 & 0.92\\
  10\% water    &          &          & 7\% $C O_2$,\ \ 8\% $H_2$,&   & \\
                &          &          & 17\% $H_2 O$&  & \\
  \hline
  water         & 1,100    & 1,960    & $H_2 O$  & 3.9 & 0.49 \\
  water         & 1,000    & 1,850    & $H_2 O$  & 3.7 & 0.46 \\
  \hline
\end{tabular}
\captionof{table}{Performance of different fuels in thermal nuclear rocket}
\end{center}

Liquid hydrogen gives by far the best performance.  One problem with using liquid hydrogen is that it has to be stored at very low temperatures.  Liquid hydrogen propellant is useful at the Asteroid Belt (far from the Sun).  Even though water has the lowest specific impulse, it can be a useful propellant within the Asteroid Belt, since it is readily available there.  The main problem with high temperature and high pressure steam applications is that such steam oxidizes boiler tubing.
There is a specific alloy, Inconel 740, containing nickel, chromium, cobalt, and no rare materials such as rhenium \cite{Inconel740}, which has outstanding resistance to steam corrosion.  When it is subjected to high pressure steam at 750$^oC$ for 10,000 hours it loses only $1.2\ mg/cm^2$ \cite{Inconel740}.

\subsection{Electrical and thermal power requirements}

\hskip.8cm In this subsection, we calculate the power requirement $P$ for a rocket engine.  We are given the engine thrust $F$, the exhaust velocity $v_0$, and engine efficiency $\eta$.  The rate of fuel consumption $(-\dot{m})$ is obtained from the fact that the force is equal to the change of momentum
\be
\label{1.03}
-\dot{m}=\frac{F}{v_0},
\ee
where the overdot denotes derivative with respect to time.
The engine expelling exhaust at the rate $(-\dot{m})$ and velocity $v_0$ supplies the power
\be
\label{1.04}
P_0=-\frac{\dot{m}\ v_0^2}{2}=\frac{F v_0}{2}
\ee
to the exhaust.  Given that the engine has efficiency $\eta$, the total power consumed by the engine is
\be
\label{1.05}
P=\frac{F v_0}{2 \eta}.
\ee
The power per unit thrust is calculated from (\ref{1.05}):
\be
\label{1.06}
p_0 \equiv \frac{P}{F}=\frac{v_0}{2 \eta}.
\ee

From (\ref{1.06}), one can see that a typical ionic engine with exhaust velocity of $50\ km/s$ and efficiency $55\%$ requires $p_0=45\ kW/N$ of electric power, where $p_0$ is power per unit thrust.  The thermal power needed to produce such electrical power is $230 - 450\ kW/N$.

Thermal nuclear propulsion uses much less energy then electrical propulsion .  First, it does not need to convert thermal energy into electrical energy.  Second, the lower exhaust velocity reduces cost per unit thrust, even though it greatly increases propellant consumption.  Finally, with some propellants thermal nuclear rocket has high efficiency.  By (\ref{1.06}), for a steam rocket with thermal efficiency $0.49$ and exhaust velocity $1,960\ m/s$, the power per unit thrust is $2,000\ W/N$.  For a 90\% methanol 10\% water rocket with thermal efficiency $0.92$ and exhaust velocity $2,860\ m/s$, the power per unit thrust is $1,560\ W/N$. Even if water tanks carried by the space transport are much heavier than the transport itself, steam rocket has much smaller power consumption then the electric rocket.  Unfortunately, water may not be available on many space routes, or $\triangle v$ may be high enough to require electrical propulsion.  Thus, in this work we will focus on energy requirements for electrical engines.

At this point we calculate power requirements for space missions.  For non--urgent space transports, an acceleration of $a_0=1 \cdot 10^{-3}\ m/s^2$ would meet most mission requirements.  Such acceleration would produce
$\triangle v =15.8\ km/s$ within six months.  Let $M_t$ be the total mass of the space ship, payload, propulsion system, propellant, and energy generation system.  Then the electric power needed for space ship propulsion can be calculated as follows:
\be
\label{1.07}
P=p_0 F= p_0 M_t a_0= M_t\ \cdot 4.5\cdot 10^4 \cdot \frac{W}{N}\ 1 \cdot 10^{-3}\ \frac{m}{s^2}=M_t\ \cdot 45\ \frac{W}{kg}.
\ee
In our work, the GCR with MHD generator has a specific electric power of
$1,000\ W/kg$.  Since the magnetoplasmodynamic thruster has specific power of $1,000\ W/kg$  as well, the whole transport has a specific power of $500\ W/kg$.  As we see from (\ref{1.07}), our transport can carry payloads 11 times its own mass, or it can deliver smaller payloads more rapidly.  For some missions, like bringing payloads from Asteroid Belt to Earth, much smaller accelerations like $a_0=2 \cdot 10^{-4}\ m/s^2$ can be used, thus much greater loads could be carried.

These power requirements can be fulfilled by space solar power as well, but the cost would be unreasonable.  In 2002, these arrays had specific power of $115\ W/kg$  and cost \$1,000 -- \$2,000 per \emph{Watt} \cite[p.432]{solar01}.  The specific power has not improved by 2012 \cite{solar02}.  For a solar power panel working 10 years or 87,660 hours, the cost of \$1,000 -- \$2,000 per $Watt$ translates to \$11.4 -- \$23 per \emph{kWh}.    Thus, nuclear fission reactors are the best sources of power for transportation in Solar System.

\section{Gas core reactor}
\subsection{General concept}

\hskip.8cm The main concern for testing new nuclear reactor types on Earth is safety.  In deep space, even if there are accidents they have no chance of contaminating human habitat.  Thus, deep space is an ideal place for testing innovative nuclear reactor ideas.

The Gas Core Reactor (GCR) is a reactor in which the fuel generates thermal energy in gaseous state.  The reactor can use either fast or thermal neutrons.  The fuel is constantly circulating between the reactor, power generating system, and the \textbf{condensing radiator}.  The \textbf{condensing radiator}, which radiates thermal energy into space is used as a heat sink.  The fuel suggested in the previous studies is uranium tetrafloride UF$_4$ \cite{UF4-01}.  Uranium tetrafloride has a melting point of 1,309$^oK$ \cite[p.39]{UF4-02} and a boiling point of 1,710$^oK$ \cite[p.87]{UF4-02}.  Uranium tetrafloride is very stable at high temperature.  At 2,500$^oK$, the equilibrium constant for the reaction
\be
\label{2.01}
UF_4 \to UF_3+F
\ee
is $1.2 \cdot 10^{-6}\ atm$.  Recall, that the equilibrium constant for the above reaction is
\be
\label{2.02}
K=\frac{[UF_3][F]}{[UF_4]},
\ee
where bracketed formulas stand for partial pressures.
The fact that UF$_3$ has very high fluorine affinity is responsible for stability and noncorrosiveness of UF$_4$.  Uranium tetrafloride is compatible with tungsten up to 3,000$^oK$ and molybdenum up to 2,000$^oK$ \cite[p.5]{UF4-01}.  In order to produce electrically conductive gas, potassium fluoride, pure potassium, rubidium fluoride and pure rubidium are added to fuel.  Potassium fluoride has melting point 1,130$^oK$ and boiling point 1,775$^oK$ \cite[p.4-76]{crc}.  Rubidium fluoride has melting point 1,106$^oK$ and boiling point 1,683$^oK$ \cite[p.4-79]{crc}.

The fuel cavity is surrounded by a molybdenum-92 shell which is 1 $cm$ to 2 $cm$ thick.\\  Molybdenum-92 has much lower neutron absorbtion cross-section then other isotopes.   The next surrounding layer consists of a neutron reflector shell which is $40\ cm$ to $60\ cm$ thick.  The best reflector is liquid beryllium metal in a pressurized molybdenum container.  Due to the fact that reflector absorbs most of the neutron energy, it will have to be cooled, thus a solid reflector such as beryllium oxide may present considerable technical problems.  The final layer is a metal pressure vessel cooled to a temperature at which it retains most of its hardness.  This vessel contains the molten beryllium reflector as well as the gas in the reactor core at 40 -- 100 $atm$.

\subsection{GCR with MHD generator}
\hskip.8cm The Liquid-Vapor Core Reactors with magnetohydrodynamic generator has been proposed by Anghaie \cite{UF4-01}.    In this reactor, a mixture of uranium tetrafluoride, potassium fluoride, and potassium acts as both the fuel and the working fluid \cite[p.5]{UF4-01}.  The fuel mixture suggested by Anghaie consists of 32\% UF$_4$ and  68\% KF by mass.  In the present work we will show that a mixture of 62\% UF$_4$, 30\% KF 8\% K by mass can perform better.  The uranium is almost pure $^{\text{235}}$U.  The fuel passes through the cycle, which is similar to the one proposed by Anghaie \cite[p.8]{UF4-01}.
Below we present the steps of the cycle which is a modification of the cycle from \cite{UF4-01}.  The modification is concerned with the modified chemical composition as well as lower temperature.  The schematic picture of GCR with MHD generator is shown on Fig. 3.
\begin{enumerate}
  \item Liquid UF$_4$ at $1500\ ^oK$ is pumped into tubes within an irradiated region of reactor surrounded by a neutron reflector.  The pumping rate is $310\ kg/s$.  The mixture is heated by fission and evaporates within the tubes.

  \item The gaseous UF$_4$ enters the reactor cavity at $310\ kg/s$.  Gaseous potassium is also pumped into reactor cavity at $40\ kg/s$.

  \item The reactor cavity is a spherical chamber surrounded by a $50\ cm$ thick neutron reflector.  The reflector is liquid beryllium within molybdenum--92 container.  Within the reactor cavity the fuel mixture is heated to $3,500\ ^oK$.  In Anghaie's model the temperature is $4,000\ ^oK$.  Using extremely high temperature will increase the electrical efficiency of the engine, but it will greatly increase corrosion.

  \item Reactor core temperature of $3,500\ ^o$K necessitates cooling of reactor walls.  One possibility of a cooling system is introduction of liquid potassium fluoride through numerous spray nozzles in the reactor walls.  KF is pumped at $150\ kg/s$.  In a similar arrangement, liquid propellant is sprayed through the nozzles in the walls of liquid propulsion rockets to shield the combustion chamber walls from the heat.

  \item The gas mixture expands through a nozzle and enters the magnetohydrodynamic generator.  The generator efficiency is 12.5\%, while in Anghaie's model it is 22\% \cite[p.4]{UF4-01}.  Our model has lower efficiency due to the lower temperature inside reactor cavity.

  \item The gas containing 62\% UF$_4$, 30\% KF 8\% K is cooled by passing through the condensing radiator, where the gas is passed through tungsten tubes which radiate energy into the space. The gas condenses into liquid.  KF can be separated from UF$_4$ since it has higher boiling point.

  \item Electric pumps pressurize the liquid to 40 -- 100 atmospheres to feed it into the reactor.

  \item On each pass a small amount of fuel may be taken to reprocessing unit, where some fission products and some fission product fluorides are removed.  Then the processed fuel is returned into the main loop.

\end{enumerate}
The whole system would generate at least 1 $kW/kg$ electrical energy.  The reactor and MHD generator are shown in the figure below:
  \begin{center}
  \includegraphics[width=11cm]{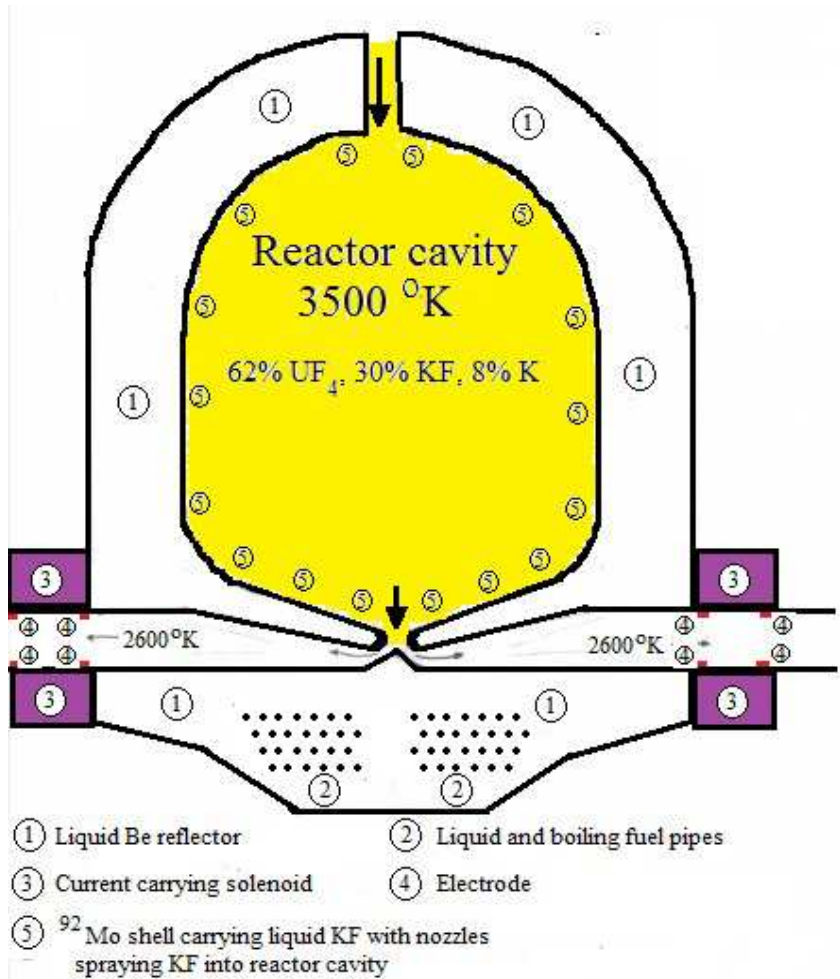}
  \vskip0cm
  \captionof{figure}{Gas core reactor with MHD generator}
  \end{center}

A Magnetohydrodynamic generator converts mechanical energy of ionized gas into electric energy.  The diagram in \cite{MHD02} illustrates its principle of operation:
  \begin{center}
  \includegraphics[width=14cm]{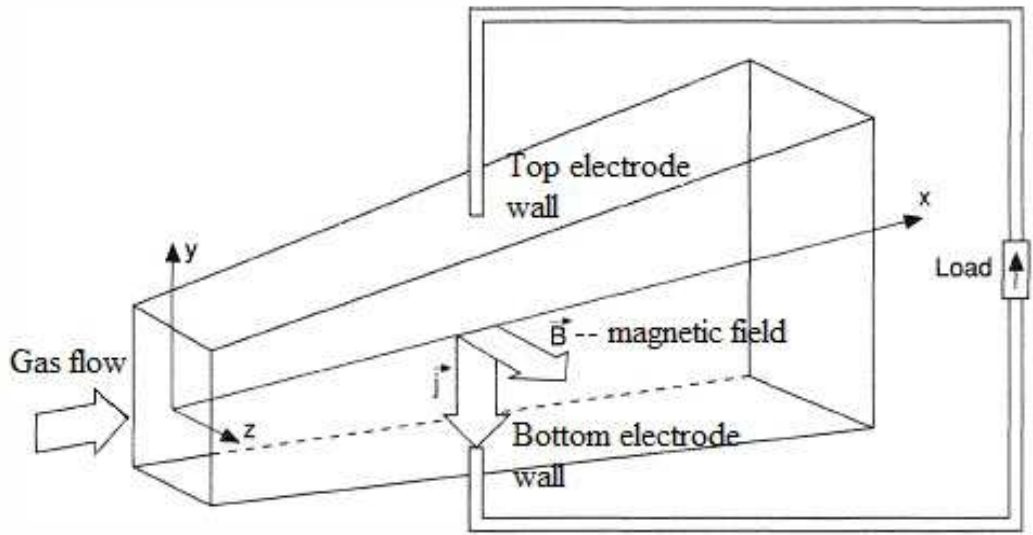}
  \vskip0cm
  \captionof{figure}{Magnetohydrodynamic generator}
  \end{center}
In the figure above the hot ionized gas flows in $x$ direction at a velocity of 1 $km/s$ to 2 $km/s$.  A strong magnetic field points in $y$ direction.  Current in $-y$ direction going through the gas is generated.
Magnetohydrodynamic generators have no moving parts making them much better suited for space missions than turbine engines.

\section{Simulation}
Neutron balance calculations are necessary to determine under which conditions the fission chain reaction is sustained.  The conditions include gaseous fuel temperature and pressure, reactor inner chamber length and diameter, moderator thickness,  inner shell thickness and material.  \emph{We show that the fission can be sustained for most reactor and moderator sizes as long as the inner shell is made out of molybdenum rather than tungsten.}
Previous works \cite{MHD03,UF4-01} constructed models of GCR for space propulsion.  Each team used a computer code for calculating neutron balance.  \emph{We have written our own Monte-Carlo simulation code in MatLab.}  The code constructs paths of a large number of neutrons within the reactor.  Path of each neutron begins with neutron generation, usually contains many collisions, and ends with neutron absorbtion or leakage from the reactor.  Using the information for  many neutrons we generate statistical data.  The data is used to determine neutron balance and criticality of the reactor.

\subsection{General method}
\hskip.8cm The reactor contains several regions with each one having its own composition.  We perform a Monte Carlo simulation tracking a number of neutrons along their paths within the reactor.  Each neutron undergoes a number of \textbf{events} before being absorbed or lost.  Below we give a list of possible events.
\begin{enumerate}
  \item An elastic or inelastic scattering during which the neutron changes both its energy and flight direction.
  \item The event of passage of the neutron from one reactor region to another, when the neutron keeps its energy and flight direction.  The reactor has three regions, they are internal cavity containing gaseous fuel, refractory material shell, and the reflector shell.
  \item Absorption or leakage of the neutron, which ends the neutron's path.  If the neutron is absorbed within a region containing fissile material, or a region where it can undergo an (n,2n) reaction, then the place of absorption along with the expected number of secondary neutrons produced are recorded.
\end{enumerate}
The neutron's energy, coordinates and flight direction are kept track of during the steps.

For a neutron traveling in a region $R=1,2,3$ (see Subsection 3.3 below) and having a given energy $E$, there are two probability constants.  The scattering probability $p_s(R,E)$ is the probability of scattering per unit length.  The absorption probability $p_a(R,E)$ is the probability of absorption per unit length.  Both probabilities have unites of $m^{-1}$.  Below we present some examples of these probabilities.  For
$1500\ ^o$K thermal neutrons in liquid beryllium reflector, $p_s=70\ m^{-1}$ and $p_a=3.4 \cdot 10^{-2}\ m^{-1}$.  For 1500$^o$K thermal neutrons in graphite reflector with density $2.0\ g/cm^3$, $p_s=47\ m^{-1}$ and $p_a=1.4 \cdot 10^{-2}\ m^{-1}$.  For 1500$^o$K thermal neutrons in $^{184}$W, $p_s=46\ m^{-1}$ and $p_a=3.8\ m^{-1}$.  As we will see in Subsection 4.2, the neutron losses within tungsten are very high.

\subsection{Tabulation of cross-sections}
\hskip.6cm The cross-sections are obtained from the South Korean source "Table of Nuclids"\cite{nuclides1} and US Government source "Evaluated Nuclear Data"\cite{nuclides2}.  A nuclide is a single isotope of a single element -- like $^{235}$U.  For each nuclide, the data is tabulated for the set of 1,231 energies from $10^{-5}\ eV$ to $20\ MeV$ given by
\be
\label{3.01}
\mathbf{E}=\left\{10^{i/100}\right\}_{i=-500}^{730}\ eV.
\ee
The data for each nuclide are presented as matrix with 1,231 rows and a few columns.  For all nonfissile nuclei, the columns contain the following:
Column 1: Energy;
Column 2: Elastic scattering cross-section;
Column 3: Radiative capture cross-section;
Column 4: First inelastic scattering cross-section;
Column N+3: N$^{\text{th}}$ inelastic scattering cross-section.
For all fissile nuclei, the following tabulation rules apply:
Column 1: Energy;
Column 2: Fission cross-section;
Column 3: Radiative capture cross-section;
Column 4: Average number of neutrons per fission.

At this point we look at the three reactor regions described in Subsection 3.3. For each reactor region the following data is tabulated for 1,231 values of neutron energy:
\begin{enumerate}
  \item EV(E) -- total event cross-section per nuclide.  An event is any scattering, absorption or fission initiation.
  \item CPT(E) -- total capture cross-section per nuclide.  Includes fission triggering.
  \item RET(E) -- the quotient of the average number of fission neutrons produced per captured neutron.
\end{enumerate}

The tables described above contain the cross-sections of every possible neutron interaction for every nuclide present within the reactor and for every possible neutron energy.  The total event cross -- section per nuclide for every reactor region can be used to find total event cross section per unit length.  It is done by the following expression:
\[
P_i(E)=n \cdot EV(E),
\]
where $P_i(E)$ is the total event cross section per unit length, and $n$ is the total nuclide density for a given reactor region.  The availability of $P_i(E)$ enables the program to simulate a neutron segment path length by (\ref{3.14}) below.  Total capture cross-section $CPT(E)$ enables the program to know the probability of each link in the neutron path being the last one.  The the quotient of the average number of fission neutrons produced per captured neutron $RET(E)$ enables the simulation program to calculate the neutron balance.

\subsection{Reactor regions}
The spherical reactor contains three regions (denoted by $R=1,2,3$) -- the gaseous fuel region, the refractory material shell, and the neutron reflector.  These regions are described below.
\begin{enumerate}
  \item \textbf{Gaseous fuel region}

  \hskip.8cm This region extends for $r \in [0,r_0]$, where $r_0$ is the radius of reactor cavity.  It contains fuel in gaseous form -- 62\% UF$_4$, 30\% KF  and 8\% K by mass.  By nuclide abundance this region consists of 10\% $^{235}$U, 28\% Potassium, and 62\% Fluorine.  An average nucleon weighs 46.2 amu.  As $^{235}$U absorbs neutrons, $^{236}$U is formed.  When $^{236}$U absorbs a neutron it produces $^{237}$U which decays to $^{237}$Np, which has high capture cross -- section for thermal neutrons. When $^{237}$Np absorbs a neutron it produces $^{238}$Np, which rapidly decays to $^{238}$Pu, which also has high thermal neutron absorption cross -- section.  When $^{238}$Pu absorbs a neutron it produces $^{239}$Pu, which is similar to $^{235}$U in its fissile properties.

  \item \textbf{Refractory material region}

  \hskip.8cm This region extends for $r \in [r_0,r_1]$. It is at most
  $2\ cm$ thick.  An obvious candidate for this region is tungsten -- it has the highest melting point and it is compatible with UF$_4$ up to 3,000$^o$K \cite[p.5]{UF4-01}.

  \hskip.8cm The isotope of tungsten with the smallest neutron absorbtion cross-section is $^{184}$ W.  Nevertheless, as it is shown in Section 4, even a very thin shell of $^{184}$ W absorbs too many neutrons for reactor to function.  Moreover, as $^{184}$ W absorbs neutrons, $^{185}$ W forms.  $^{185}$ W may absorb a neutron to form $^{186}$ W, or it may decay to $^{185}$ Re -- it's $\beta$ -- decay half -- life is 75 days.  Further neutron absorptions and $\beta$ -- decays will produce $^{187}$ Re, osmium isotopes 186 -- 190, as well as iridium.  All rhenium, osmium and iridium isotopes have high thermal neutron capture cross -- sections.

  \hskip.8cm According to Anghaie, "for reasons of neutron economy, tungsten and tungsten alloys cannot be used. Instead, materials such as the alloy TZM (99\% Mo, 0.9\% Ti, 0.1\% Zr) are necessitated both from a materials and a neutronics" \cite[p.5]{UF4-01}.  Molybdenum used in this region will consist of isotope $^{92}$Mo.

  \item \textbf{Reflector region}

  \hskip.8cm This region extends for $r \in [r_1,r_2]$.  It is is
  $40\ cm$ to $60\ cm$ thick.  It contains a neutron reflector for which there are several choices: i) molten beryllium in a container made of molybdenum isotope $^{92}$Mo, ii) beryllium oxide, iii) graphite,  iv) molten beryllium fluoride BeF$_2$ in a container made of molybdenum isotope $^{92}$Mo.  In our simulation we use molten beryllium.  As $^9$Be absorbs neutrons, the isotope $^{10}$Be is produced.  The isotope $^{10}$Be is metastable with half-life of $1.5\cdot 10^6\ years$ \cite[p. 11-199]{crc} and neutronic properties similar to those of $^9$Be.

  \hskip.8cm A small amount of $^9$Be will undergo the reaction
  \be
  \label{3.02}
  ^9Be+n \to ^6Li+^4He.
  \ee
  The isotope $^6$Li has a thermal neutron absorbtion cross--section of
  $920\ barn$, thus beryllium would have to be constantly reprocessed to separate lithium.  Given that beryllium is liquid it should not be difficult.

\end{enumerate}

\subsection{Path of a single neutron}
\hskip.8cm The simulation will generate the paths of $N_n$ neutrons each of which originates from the gaseous fuel region of the reactor.  The energy distribution of neutrons from $^{235}$U fission is \cite[p. 34]{nenergy}:
\be
\label{3.03}
N(E)=0.453 \exp \left(-1.036 E \right)
\sinh \left( \sqrt{2.29 E} \right),
\ee
where the energy $E$ is given in $MeV$.

Each path will contain one or more segments described below.
\begin{center}
  \textbf{Neutron path segment}
\end{center}

We define any segment of the neutron path by the following parameters.  $\textbf{r}_s \in \mathbb{R}^3$ is the origin of the segment.  The velocity direction $\textbf{u}_s$ belongs to a unit sphere.  The energy of the neutron is given in $eV$, and determines absorption, scattering and fission cross -- sections and thus absorption probabilities for the neutron.  The infinite line $\mathbb{L}_s \subset \mathbb{R}^3$ is the linear extension of the neutron path segment in both directions.  We express $\mathbb{L}_s$ as
\be
\label{3.04}
\mathbb{L}_s=\left\{
\textbf{r}_s+x \textbf{u}_s,\ \ x \in \mathbb{R}
\right\}.
\ee
The neutron path segment is illustrated below:
  \begin{center}
  \includegraphics[width=4cm]{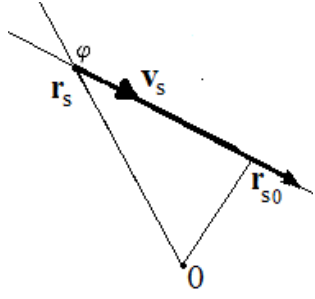}
  \vskip0cm
  \captionof{figure}{Neutron path segment}
  \end{center}
The angle between $\textbf{r}_s$ and $\textbf{u}_s$ is $\varphi$ determined by
\be
\label{3.05}
\cos \varphi= \frac{\textbf{r}_s \cdot \textbf{u}_s}
{\|\textbf{r}_s\| \|\textbf{u}_s\|}=
\frac{\textbf{r}_s \cdot \textbf{u}_s}
{\|\textbf{r}_s\| }.
\ee
The point at which $\mathbb{L}_s$ is closest to the origin is given by
\be
\label{3.06}
\textbf{r}_{s0}=\textbf{r}_s-
\|\textbf{r}_s\|\left( \cos \varphi \right) \textbf{u}_s=
\textbf{r}_s- \left(\textbf{r}_s \cdot \textbf{u}_s\right) \textbf{u}_s.
\ee
The closest approach of $\mathbb{L}_s$ to the origin is
\be
\label{3.07}
\begin{split}
\left\|\textbf{r}_{s0}\right\|&=
\left\|
\textbf{r}_s- \left(\textbf{r}_s \cdot \textbf{u}_s\right) \textbf{u}_s
\right\|=
\sqrt{
\left(
\textbf{r}_s- \left(\textbf{r}_s \cdot \textbf{u}_s\right) \textbf{u}_s
\right)
\cdot
\left(
\textbf{r}_s- \left(\textbf{r}_s \cdot \textbf{u}_s\right) \textbf{u}_s
\right)
}
=\sqrt{
\| \textbf{r}_s \|^2 -\left(\textbf{r}_s \cdot \textbf{u}_s\right)^2
}.
\end{split}
\ee
Notice, that if $\cos \varphi<0$, then the point of closest approach
$\textbf{r}_{s0}$ is in the direction of the neutron path, otherwise it is virtual.

At this point we calculate the value of $x$ for which the neutron path intersects the sphere centered at the origin with radius $r_i$.  We have to solve the equation
\be
\label{3.08}
\left\| \textbf{r}_s+x \textbf{u}_s \right\|=r_i.
\ee
Rewrite (\ref{3.08}) as
\be
\label{3.09}
\begin{split}
r_i^2&=\left\| \textbf{r}_s+x \textbf{u}_s \right\|^2
=
\textbf{r}_s \cdot \textbf{r}_s+
2 \textbf{u}_s \cdot \textbf{r}_s x+
\textbf{u}_s \cdot \textbf{u}_s x^2
=x^2+ 2 \textbf{u}_s \cdot \textbf{r}_s x+\textbf{r}_s \cdot \textbf{r}_s,
\end{split}
\ee
which can be rewritten as a quadratic equation
\be
\label{3.10}
\begin{split}
x^2+ 2 \textbf{u}_s \cdot \textbf{r}_s x+\|\textbf{r}_s\|^2-r_i^2=0.
\end{split}
\ee
Thus,
\be
\label{3.11}
\begin{split}
x_s=-\textbf{u}_s \cdot \textbf{r}_s
\pm
\sqrt{
\left( \textbf{u}_s \cdot \textbf{r}_s \right)^2
-\|\textbf{r}_s\|^2+ r_i^2.
}
\end{split}
\ee
Notice, that the intersection between the neutron path and the sphere of radius $r_i$ is possible only if two conditions are satisfied.  First, $x_s>0$.  Second, no other sphere where reactor regions are changing is crossed for an $x<x_s$.  Thus, if (\ref{3.11}) has two positive solutions, we take only the smaller one.  Also notice, that for any point within the reactor and any direction of the neutron path there is a value $x_s$, which is the smallest positive value for which the neutron changes reactor region or escapes the reactor.

If the neutron continues its flight in a straight line without undergoing a collision or absorption event, then the flight segment $\mathbb{S}_s$ would be given by
\be
\label{3.12}
\mathbb{S}_s=\left\{
\textbf{r}_s+x \textbf{u}_s,\ \
x \in [0,x_s]
\right\}.
\ee
Let $P_i(E)$ be the probability per unit length of the neutron with energy $E$ undergoing a scattering, absorption, or triggering a fission in the reactor region $i$.  The probability of an event along $\mathbb{S}_s$ is
\be
\label{3.13}
P=1-\exp \left(-\frac{x_s}{P_i(E)} \right).
\ee

In a simulation, the value $x_e$ where the neutron either changes the reactor regions or undergoes scattering, absorption, or triggering a fission can be obtained in two steps.  First choose $u \in [0,1]$ from a uniform random distribution.  Then, set $x_e$ as
\be
\label{3.14}
x_e=\min \left\{x_s,P_i(E) \ln u \right\}.
\ee
A new segment starts from the point
\be
\label{3.15}
\textbf{r}_{s+1}=\textbf{r}_s+x_e \textbf{u}_s.
\ee
\begin{center}
  \textbf{Neutron Scattering}
\end{center}

Suppose a neutron of mass $m_0$, flight direction $\textbf{u}_0$ and energy $E_0$ is scattered by a nucleus of mass $m_1$, flight direction $\textbf{u}_1$ and energy $E_1$.  The velocity of the neutron is
\be
\label{3.16}
\textbf{v}_0=\sqrt{\frac{2 E_0}{m_0}}\ \textbf{u}_0.
\ee
We do not know the energy of the scattering nucleus, thus we will have to generate a random energy.  We do have the distribution of particles at temperature $T$ given by \cite[p.312]{Boltzmann}:
\be
\label{3.17}
P(E)=C \sqrt{E} \exp \left( \frac{E}{k T} \right),
\ee
where $k=1.381 \cdot 10^{-23}\ J/^oK$ is the Boltzmann constant.

The velocity of the nucleus is
\be
\label{3.18}
\textbf{v}_1=\sqrt{\frac{2 E_1}{m_1}}\ \textbf{u}_1.
\ee
The velocity of the center of mass is
\be
\label{3.19}
\textbf{v}_c=\frac{m_0 \textbf{v}_0+m_1 \textbf{v}_1}{m_0+m_1}.
\ee
The velocity of the neutron relative to the center of mass is
\be
\label{3.20}
\begin{split}
\textbf{v}_{0c}&=\textbf{v}_{0}-\textbf{v}_{c}=
\textbf{v}_{0}-\frac{m_0 \textbf{v}_0+m_1 \textbf{v}_1}{m_0+m_1}=
\frac{m_1 \left(\textbf{v}_0-\textbf{v}_1\right)}{m_0+m_1}.
\end{split}
\ee
The speed of the neutron relative to the center of mass is
\be
\label{3.21}
\begin{split}
v_{0c}=
\frac{m_1 \left\| \textbf{v}_0-\textbf{v}_1\right\| }{m_0+m_1}.
\end{split}
\ee
The velocity of the nucleolus relative to the center of mass is derived similar to (\ref{3.20}):
\be
\label{3.22}
\begin{split}
\textbf{v}_{1c}=
\frac{m_0 \left(\textbf{v}_1-\textbf{v}_0\right)}{m_0+m_1}.
\end{split}
\ee
At this point we consider three types of scattering: i) Low energy elastic; ii) High energy elastic; iii) Inelastic.
\begin{enumerate}
  \item \textbf{Low energy elastic scattering}

  \hskip.8cm For energies under $50\ keV$, the angular distribution of scattering neutrons is spherically symmetric. According to "Evaluated Nuclear Data"\cite{nuclides2}, for light scattering nuclei up to oxygen and neutron energies up to $50\ keV$, the angular distribution of scattered neutrons does not deviate from spherical symmetry by more then 1\%.  In this case, the scattering direction $\textbf{u}_0^*$ is selected from a spherical uniform distribution via the following set of two commands in MatLab
  \begin{verbatim}
    uu=randn(1,3);
    uu=uu/norm(uu);
  \end{verbatim}
  \vskip-0.8cm
  The neutron speed relative to the center of mass is given by (\ref{3.21}).  In an elastic collision, the neutron will keep it's speed relative to the center of mass,  but change it's direction.  Thus the new neutron velocity relative to the center of mass is
  \be
 \label{3.23}
 \begin{split}
 \textbf{v}_{0c}^*=
 \frac{m_1 \left\| \textbf{v}_0-\textbf{v}_1\right\| }{m_0+m_1} \textbf{u}_0^*.
 \end{split}
 \ee
 Adding the center of mass velocity to the neutron's velocity we obtain the scattered neutron's velocity:
   \be
 \label{3.24}
 \begin{split}
 \textbf{v}_{0}^*=\textbf{v}_{0c}^*+\textbf{v}_c=
 \frac{m_1 \left\|\textbf{v}_0-\textbf{v}_1\right\|}{m_0+m_1}\textbf{u}_0^*
 +
 \frac{m_0 \textbf{v}_0+m_1 \textbf{v}_1}{m_0+m_1}.
 \end{split}
 \ee
 \item \textbf{High energy elastic scattering}

 \hskip.8cm Once we know the scattering direction, the equations (\ref{3.23}) and (\ref{3.24}) are the same as for the low energy scattering.  The main difference is that the scattering direction $\textbf{u}_0^*$ is neither spherically symmetric nor independent of the original direction $\textbf{u}_0$.  Let $\beta$ be the scattering angle which is the angle between $\textbf{u}_0^*$ and $\textbf{u}_0$.  Then
 \be
 \label{3.25}
 \cos \beta= \textbf{u}_0^* \cdot \textbf{u}_0.
 \ee
 For symmetrical scattering, $\cos \beta$ follows a uniform random distribution within $[-1,1]$ and thus the expectation value of
 $\cos \beta$ is $0$:
  \be
 \label{3.26}
 E(\cos \beta)=0.
 \ee
 For high energy neutrons, and especially for heavier nuclei, the scattering is cylindrically symmetric, but not spherically symmetric.  "Evaluated Nuclear Data"\cite{nuclides2} tabulates the distributions of $\cos \beta$ for all scattering nuclides for different incoming neutron energies.

 \hskip.8cm In practical terms, the expected value of $\cos \beta$ increases with neutron energy and approaches $1$ as the neutron energy comes into the region of hundreds of MeV -- although such energetic neutrons are not considered in the present work.  That means that the deflection angle is smaller for neutrons with energies of over $2\ MeV$ than for thermal neutrons.  Such neutrons produced within the reactor core will travel deeper into the neutron reflector before being stopped than the less energetic neutrons.  Thus, those of them which will be neither absorbed nor lost will be thermalized before returning to the reactor core.

  \item \textbf{Inelastic scattering}

  \hskip.8cm  In inelastic scattering, part of the neutron's initial energy is transferred to the scattering nucleus, which goes into an excited state.  The excitation energy $E_e$ of the nuclide has to be lower than the energy of the neutron -- nucleus system with respect to the center of mass, otherwise known as the center of mass energy.  The excitation energy in an inelastic scattering is taken from the system energy relative to the center of mass.

  \hskip.8cm In order to determine the energy of the system relative to the center of mass, we recall the neutron and nucleus velocity in the center of mass frame of reference:
  \be
  \label{3.27}
  \begin{split}
  v_{0c}=
  \frac{m_1 \left\| \textbf{v}_0-\textbf{v}_1\right\| }{m_0+m_1},
  \qquad
    v_{1c}=
  \frac{m_0 \left\| \textbf{v}_0-\textbf{v}_1\right\| }{m_0+m_1}.
  \end{split}
  \ee
  Given that the neutron energy of inelastic scattering is at least $50\ keV$ for all nuclei considered, the velocity of the nucleus is negligible with respect to neutron velocity.  Thus,
  \be
  \label{3.28}
  \begin{split}
  v_{0c}=
  \frac{m_1  }{m_0+m_1}v_0,
  \qquad
    v_{1c}=
  \frac{m_0 }{m_0+m_1}v_0,
  \end{split}
  \ee
  where $v_0=\|\textbf{v}_0\|$ is the speed of the neutron in the reactor frame of reference.  The total kinetic energy of the neutron and the nucleus in the center of mass frame of reference is
  \be
  \label{3.29}
  \begin{split}
  E_c&=\frac{1}{2}\left(m_0 v_{0c}^2+m_1 v_{0c}^2 \right)=
  \frac{1}{2}\left[
  m_0 \left(\frac{m_1  }{m_0+m_1}v_0 \right)^2+
  m_1 \left( \frac{m_0 }{m_0+m_1}v_0 \right)^2
  \right]\\
  &=\frac{1}{2} m_0 v_0^2
  \left[
  \frac{m_1^2}{\left(m_0+m_1\right)^2}+
  \frac{m_1 m_0}{\left(m_0+m_1\right)^2}
  \right]=E_0 \frac{m_1}{m_0+m_1}.
  \end{split}
  \ee
  After the scattering event, the center of mass kinetic energy will be
  \be
  \label{3.30}
  E_c^*=\frac{m_1 E_0}{m_0+m_1}-E_e.
  \ee
  As a result of the inelasticity, the center of mass energy is decreased by the factor of
  \be
  \label{3.31}
  \frac{E_c^*}{E_c}=
  \frac{\frac{m_1 E_0}{m_0+m_1}-E_e}
       {\frac{m_1 E_0}{m_0+m_1}}=
  1-\frac{E_e}{E_c} \frac{m_0+m_1}{m_1}.
  \ee
  Therefore, in the inertial frame of reference, both the speed of the scattered neutron and the scattered nuclei will decrease by the factor of
    \be
  \label{3.32}
  \sqrt{1-\frac{E_e}{E_c} \frac{m_0+m_1}{m_1}}.
  \ee
  Substituting (\ref{3.32}) into (\ref{3.23}), we obtain the velocity of an inelastically scattered neutron relative to the center of mass:
  \be
 \label{3.33}
 \begin{split}
 \textbf{v}_{0c}^*=
 \frac{m_1 \left\| \textbf{v}_0-\textbf{v}_1\right\| }{m_0+m_1}
 \sqrt{1-\frac{E_e}{E_c} \frac{m_0+m_1}{m_1}}\
 \textbf{u}_0^*.
 \end{split}
 \ee
   Adding the center of mass velocity to the neutron's velocity we obtain the scattered neutron's velocity:
   \be
 \label{3.34}
 \begin{split}
 \textbf{v}_{0}^*=\textbf{v}_{0c}^*+\textbf{v}_c=
 \frac{m_1 \left\|\textbf{v}_0-\textbf{v}_1\right\|}{m_0+m_1}
 \sqrt{1-\frac{E_e}{E_c} \frac{m_0+m_1}{m_1}}\
 \textbf{u}_0^*
 +
 \frac{m_0 \textbf{v}_0+m_1 \textbf{v}_1}{m_0+m_1}.
 \end{split}
 \ee
\end{enumerate}

\subsection{Neutronics}

\hskip.8cm Neutronics refers to the balance of neutrons which has to be maintained within reactor.  We introduce \textbf{criticality}, otherwise known as multiplication factor, which is the ratio of the number of neutrons produced by fission to the number of neutrons absorbed or lost \cite[p.3]{neutronicsA}:
\be
\label{3.35}
k=\frac{\text{neutrons produced in current fission generation}}
{\text{neutrons absorbed in previous fission generation}} .
\ee
If $k<1$, the reactor is \textbf{subcritical} and fission stops very quickly.  If $k>1$, the reactor becomes supercritical and the neutron flux multiplies until the criticality is restored to 1 or there is an accident.  Thus for a working reactor, $k=1$ with possible oscillations of very small magnitude.

Define the \textbf{fuel utilisation factor} \cite[p.4]{neutronicsA} as
\be
\label{3.36}
f=\frac{\text{number of neutrons absorbed by the fuel}}
{\left\{\begin{split}
&\text{number of neutrons absorbed by the fuel,  }\\
&\text{refractory sphere, reflector, fission products,} \\
&\text{ and leaked neutrons}
 \end{split} \right\} }.
\ee
Define \textbf{regeneration factor} \cite[p.4]{neutronicsA} as
\be
\label{3.37}
\eta=\frac{\text{neutrons produced from fission}}
{\text{neutrons absorbed in fuel}} .
\ee
As we will see in simulations, for the reactor we are considering,
$\eta=1.96$.  The relationship between criticality, fuel utilisation factor and regeneration factor is
\be
\label{3.38}
k=f \eta =1.
\ee
It follows, that the proportion of the neutron flux absorbed by the fuel is
\be
\label{3.39}
f=\frac{1}{\eta}=51\%.
\ee
The total fraction of neutrons leaked from  reactor, absorbed by fission products, the $^{92}$Mo shell, and reflector and lost in other ways has to be 49\%.

In the gas core reactor, the density of fuel will be much lower than that of solid or liquid core reactors.  Thus, ensuring that 51\% of the neutrons are absorbed by the fuel would be a challenge.  In order to do so we need the following:
\begin{enumerate}
  \item Maximize the fuel density. To do so we maximize the pressure within the limits presented by the pressure vessel.
  \item Maximize the size of reactor cavity within the limits imposed by reactor mass.  Even though reactors existing on Earth have almost no mass limit, the cost of placing material into deep space is at least \$20,000 per kg.
  \item Use reflector of sufficient thickness to minimize neutron leakage.  Generally, $40\ cm$ to $60\ cm$ reflectors will be used.

  \item Use the least amount of tungsten or no tungsten at all in the refractory shell.  Tungsten is compatible with UF$_4$ vapor up to 3,000$^o$K, while molybdenum is compatible with UF$_4$ vapor up to 2,000$^o$K only \cite[p.5]{UF4-01}.

  \item Separate fission products with high neutron absorbtion cross -- section from the fuel stream.  This should be possible, since the fuel is constantly recirculating in and out of reactor zone.  These products can be discarded into space.

      Most metal fission products will appear as fluorides.  Products with boiling point significantly different from that of uranium tetrafluoride (1417 $^o$C) can be easily separated by distillation.  The worst neutron absorber $^{135}$Xe cannot be separated as it absorbs neutrons very fast.

  \item Discard the fuel as it is strongly contaminated by $^{236}$U.  This will mean that some $^{235}$U will also be discarded -- in our model 60\%.
\end{enumerate}
\begin{center}
  \textbf{Neutron absorbtion by fission products}
\end{center}

\hskip.5cm As $^{235}$U undergoes fission, fission products will form.  Metallic products will be present mostly in the form of fluorides. Neutron absorption by fission products is called poisoning \cite{neutronics}.  Some important neutron absorbers or \textbf{neutron poisons} are presented in Table 3.  \textbf{Fission yield} denotes the number of nuclei produced per fission.
  \begin{center}
    \begin{tabular}{|l|l|l|l|}
       \hline
       Isotope & Fisson & Thermal neutron&  Comment \\
               & yield  & cross section &           \\
               &        & at $0.025\ eV$&           \\
       \hline
       $^{135}$Xe & 0.065 & $2.6 \cdot 10^6$ barn& Absorbs neutrons very fast \\
       \hline
       $^{149}$Sm & 0.011 & $4.0 \cdot 10^4$ barn & Absorbs neutrons fast, produces $^{150}$Sm \\
                  &       &                  &   with 109 barn thermal neutron \\
                  &       &                  &   absorption cross section \\
       \hline
       $^{151}$Sm & 0.008  & $1.5 \cdot 10^4$ barn  & Absorbs neutrons fast, produces $^{152}$Sm     \\
       & & & with 2,760 barn neutron resonance integral \\
       \hline
     \end{tabular}
     \captionof{table}{Important neutron poisons}
  \end{center}

According to Kotov \cite[p. 190]{Kotov}, the total fission product absorbtion (FPA)  is equivalent to  $FPA=0.11$ per each fission event.  In order to account for fission product absorbtion, we multiply the number of neutrons produced per fission by $(1-FPA)$.  V. Kotov has developed a regime for which $FPA=0.07$.  In this regime the fuel is used in a cycle for 5 to 10 hours and then is rested for 35 -- 50 hours during which most $^{135}$Xe decays \cite[p. 191]{Kotov}.  Given that fuel can be stored in liquid form, possibly this cycle can be used.
\begin{center}
  \textbf{Neutron multiplication by beryllium}
\end{center}

Neutrons are multiplied in beryllium via the reaction
\be
\label{3.40}
^9\text{Be} +n \to ^8\text{Be}+2 n -1.665\ MeV\to 2 ^4 \text{He}+2 n.
\ee
This reaction is possible only for neutrons with energy exceeding 1.8 $MeV$.  The scattered neutrons have much lower energy than the original neutron. Only the most energetic neutrons can produce tertiary neutrons.  According to a 1959 calculation \cite{BeMult01}, beryllium moderator increases the number of neutrons by 5.1\% to 7.9\%.  According to a 1963 Soviet calculation \cite{BeMult02}, number of neutrons is increased by 10\% in beryllium and 8\% in beryllium oxide.  In our calculations, the number of neutrons is increased by 9\% to 11\%.  For beryllium moderator, the neutron gain almost offsets the absorbtion loss.

\section{Simulation Results}
In this section we present the results of numerical simulations concerning neutron balance.  These simulations show that the reactor can not achieve criticality if the shell surrounding reactor cavity is made of tungsten over 1 $mm$ thick.  The reactor can easily achieve criticality if it has a $^{\text{92}}$Mo shell 1 $cm$ thick.

\subsection{Neutron reflector}

\hskip.5cm At this point we determine the optimal thickness of neutron reflector.  To this end we look at the paths of the neutrons which are reflected by liquid beryllium shell and returned into the reactor cavity.  Let $p(x)$ be a fraction of reflected neutrons which have achieved depth of at least $x$ within the neutron reflector.  In all cases, the function $p(x)$ is strictly decreasing and such that $p(0)=1$, and
\be
\label{4.01}
\lim_{x \to \infty} p(x)=0.
\ee
Let $R(r)$ be the neutron reflectance of an infinite liquid beryllium shell surrounding a reactor cavity of radius $r$.  Then a liquid beryllium shell of thickness $x$ surrounding a reactor cavity of radius $r$ would have reflectance $(1-p(x)) R(r)$.

\begin{center}
  \textbf{Liquid beryllium reflector}
\end{center}

We estimate the function $p(x)$ for liquid beryllium reflectors with reactor cavities of different radii. In each case, we have used 20,000 neutrons.  For these simulations, we have used the programs \emph{InfiniteCilynderReflector\_Be.m} and \emph{SphericalReflector\_Be.m} which we have written ourselves.  The results are tabulated below:
\begin{center}
 \begin{tabular}{|p{1.5cm}|c|c|c|c|c|c|c|c|c|c|c|c|}
   \hline
   Reactor cavity & & \multicolumn{9}{|c|}{Penetration depth $x$, cm} \\
   radius $cm$     & & 30 & 35 & 40 & 45 & 50 & 55 & 60 & 65 & 70 \\
   \hline
   50            &$p(x)=$ & 3.5\% & 2.3\% & 1.7\%& 1.1\%& 0.8\% & 0.6\% &0.36\% & 0.27\% &0.17\% \\
   70            &$p(x)=$ & 3.7\% & 2.6\% & 1.9\%& 1.4\%& 1.0\% & 0.7\% &0.48\% & 0.33\% &0.17\% \\
   100            &$p(x)=$ & 3.3\% & 2.3\% & 1.5\%& 1.1\%& 0.8\% & 0.5\% &0.35\% & 0.20\% &0.14\% \\
   150            &$p(x)=$ & 3.3\% & 2.2\% & 1.5\%& 1.1\%& 0.7\%& 0.5\% & 0.38\% & 0.26\% & 0.18\% \\
   200            &$p(x)=$ & 3.5\% & 2.3\% & 1.6\% & 1.2\% & 0.8\% & 0.5\%& 0.39\% & 0.30\% & 0.25\%\\
   \hline
 \end{tabular}
 \captionof{table}{Reflected neutrons penetrating liquid beryllium reflector to depth $x$}
\end{center}
As it will be seen from Table 5 below, the \emph{optimal thickness} of molten beryllium neutron reflector should be $40\ cm$ to $60\ cm$ for most reactors.

The reflectance $R(r)$ has also been estimated for different radii of reactor cavity.  We have also performed several tests for reactors shaped as infinite cylinders rather than spheres.  In each case, the simulation ran a set of 20,000 neutrons.  For all simulations, we have used the programs \emph{InfiniteCilynderReflector\_Be.m} and \emph{SphericalReflector\_Be.m} mentioned above.  The reflectances are tabulated below:
\begin{center}
  \begin{tabular}{|l|l|l|l|l|}
    \hline
    Cavity     & Neutron         & Neutron  & Neutron         & Neutron          \\
    radius $r$ & Reflectance $R$ & Reflectance $R$ & Reflectance $R$ & Reflectance $R$ \\
    in $cm$    & for fission     & for thermal  & for fission     & for thermal    \\
               & neutrons        & neutrons    & neutrons        & neutrons     \\
               & spherical       & spherical  & cylindrical      &  cylindrical      \\
               & reactor         & reactor   & reactor         & reactor        \\
    \hline
     50 & $85.9 \pm 0.2 \%$ & $91.6 \pm 0.2\%$ & $88.5 \pm 0.2\%$ &
     $93.4 \pm 0.2\%$\\
     70 & $87.1 \pm 0.2 \%$ & $92.6 \pm 0.2\%$ & $88.9 \pm 0.2\%$ &
     $93.8 \pm 0.2\%$\\
    100 & $88.6 \pm 0.2 \%$ & $93.3 \pm 0.2\%$ & $90.0 \pm 0.2\%$ &
    $94.2 \pm 0.2\%$\\
    150 & $89.8 \pm 0.2 \%$ & $93.6 \pm 0.2\%$ & $90.8 \pm 0.2\%$ &
    $94.4 \pm 0.2\%$\\
    200 & $90.0 \pm 0.2 \%$ & $93.7 \pm 0.2\%$ & $90.8 \pm 0.2\%$ &
    $94.6 \pm 0.2\%$\\
    \hline
  \end{tabular}
  \captionof{table}{Infinite beryllium sphere or cylinder neutron reflectance}
\end{center}
As follows from Table 5, reflectance $R$ for fission neutrons in a spherical reactor is a little bit lower than the reflectance for fission neutrons in a cylindrical reactor.  A similar conclusion can be made about the reflectance for thermal neutrons.

In operation of an actual reactor, a fission neutron is usually reflected several times before being lost or absorbed.  As we will see below,  27\% to 36\% of neutrons will be lost or absorbed by the reflector.  For the reactors with criticality of 1, reflector loss would be 31\% to 33\%.  Thus, an average neutron is reflected about three times.

\subsection{Reactor neutron balance}

\hskip.8cm We have performed several simulations for a spherical reactor surrounded by a neutron reflector shell.  Here we make calculations for a reactor with cavity surrounded by a tungsten shell.  The tungsten shell initially contains $^{184}$W.  Eventually, as $^{184}$W absorbs neutrons it will contain $^{185}$W, $^{185}$Re, $^{186}$Re, $^{187}$Re, $^{188}$Re,
$^{186}$Os, $^{187}$Os, $^{188}$Os, $^{189}$Os, $^{190}$Os, and iridium.  All of these isotopes have large thermal and resonance neutron absorbtion cross-sections.  It is convenient to introduce the \textbf{effective thickness} of tungsten shell which is thickness of non-irradiated shell of $^{184}$W, which has the same neutron absorbtion properties as an irradiated $^{184}$W shell.  Depending on irradiation, the effective thickness of $^{184}$W shell should be up to 4 times greater than the actual thickness of the $^{184}$W shell.

We have performed tests for reactors with 50 $cm$ molten beryllium reflector shells.  We use reactors with cavity radii of 1.0 $m$ and 1.5 $m$.  We use fuel gas densities of 20 $kg/m^3$ and 40 $kg/m^3$.
We use tungsten shells of effective thickness of 0 $mm$, 1 $mm$, 2 $mm$, and 5 $mm$.  In each case the sample size of 5,000 neutrons have been used.  In all working reactors, the multiplication factor $k$ is 1.  If our calculations show that $k>1$, it means that there is room for relaxing reactor conditions -- like reducing fuel density, or reducing reflector thickness, or using a thicker tungsten shell.  If our calculations show that $k<1$, it means that the reactor will not work until adjustments are made.  For all simulations, we have used the program \emph{SphericalReactor\_Be.m} which we have written ourselves.
The program calculates average number of fission neutrons produced per neutron absorbed by fuel.  This number is always $1.90$ to $1.96$.  By multiplying this number by the fraction of the neutrons absorbed by fuel we obtain the multiplication factor.
The results are tabulated below.
\setcounter{table}{4}
\vskip.8cm
\begin{center}
\begin{longtable}{|p{1.2cm}|p{1.6cm}|p{1.7cm}|p{1.2cm}|p{1.6cm}|p{1.6cm}|p{1.6cm}|p{1.3cm}|p{1.0cm}|}
\hline
Reactor cavity radius $cm$ &
Tungsten shell effective thickness $mm$ &
Beryllium reflector thickness&
Fuel density $kg/m^3$ &
Percent neutrons absorbed by fuel&
Percent neutrons absorbed by W&
Percent neutrons absorbed by reflector&
Percent neutrons lost&
Multi- plication factor\\
\hline
100 & 0  & 50 & 20 & 59.1 & 0  & 20.8 & 20.1  & 1.15  \\ 
100 & 0  & 50 & 40 & 64.4 & 0  & 18.5 & 17.1  & 1.23  \\ 
150 & 0  & 50 & 20 & 65.0 & 0  & 18.2 & 16.8  & 1.24  \\ 
150 & 0  & 50 & 40 & 67.5 & 0  & 17.1 & 15.4  & 1.28  \\ 
\hline
100 & 1  & 50 & 20 & 49.5 & 9.1 & 21.3 & 20.1 & 0.97  \\ 
100 & 1  & 50 & 40 & 56.6 & 6.8 & 18.8 & 17.8 & 1.09  \\ 
150 & 1  & 50 & 20 & 55.6 & 8.0 & 19.5 & 16.9 & 1.07  \\ 
150 & 1  & 50 & 40 & 59.6 & 7.2 & 18.3 & 14.9 & 1.14  \\ 
\hline
100 & 2  & 50 & 20 & 47.8 & 12.9 & 19.8 & 19.5 & 0.93 \\
100 & 2  & 50 & 40 & 52.6 & 11.5 & 19.0 & 16.9 & 1.01 \\
150 & 2  & 50 & 20 & 53.7 & 11.7 & 18.0 & 16.7 & 1.03 \\ 
150 & 2  & 50 & 40 & 58.1 & 10.6 & 15.6 & 15.8 & 1.11 \\ 
\hline
100 & 5  & 50 & 20 & 39.4 & 22.5 & 19.1 & 19.1 & 0.77 \\ 
100 & 5  & 50 & 40 & 45.1 & 19.6 & 17.8 & 17.5 & 0.87 \\
150 & 5  & 50 & 20 & 46.4 & 20.7 & 17.4 & 15.5 & 0.90 \\ 
150 & 5  & 50 & 40 & 50.6 & 17.8 & 16.3 & 15.3 & 0.96 \\ 
\hline
\end{longtable}
\captionof{table}{Neutron balance for a reactor with beryllium reflector and tungsten inner shell }
\end{center}

The above table shows that a tungsten shell of even 1 $mm$ effective thickness absorbs 7\% -- 9\% of neutrons.
(See ``Percent neutrons absorbed by W" column and take data for Tungsten shell effective thickness equal to one.)
The neutrons absorbed by the tungsten shell are almost completely reflected neutrons, thus a 1 $mm$ effective thickness lowers criticality by .14 -- .18.
These numbers are obtained by subtracting the multiplication factors for reactors with 1 $mm$ tungsten shell from multiplication factors for similar reactors with no tungsten shell.
Given that even a 0.5 $mm$ tungsten shell will soon attain an effective thickness of 2 $mm$, it follows, that very little tungsten can be used as the inside shield.  For lighter reactors, no tungsten at all should be used for the internal shell.

Instead of tungsten, an alloy consisting almost entirely of molybdenum -- 92 is used.  Once again, we perform tests for reactors with with 50 $cm$ molten beryllium reflector shells.  We use reactors with cavity radii of 1.0 $m$ and 1.5 $m$.  We use fuel gas densities of 20 $kg/m^3$ and 40 $kg/m^3$.
We use $^{92}$Mo shells of effective thickness of 1 $cm$ and 2 $cm$.   The molybdenum shell should be thick since it contains many tubes carrying potassium and potassium fluoride.  Molten potassium and potassium fluoride are sprayed from many micronozzles on the reactor walls, thus shielding the walls from the extreme heat.

\begin{center}
\setcounter{table}{5}
\begin{longtable}{|p{1.2cm}|p{1.6cm}|p{1.7cm}|p{1.2cm}|p{1.6cm}|p{1.6cm}|p{1.6cm}|p{1.3cm}|p{1.0cm}|}
\hline
Reactor cavity radius $cm$ &
$^{92}$Mo shell effective thickness $cm$ &
Beryllium reflector thickness&
Fuel density $kg/m^3$ &
Percent neutrons absorbed by fuel&
Percent neutrons absorbed by Mo&
Percent neutrons absorbed by reflector&
Percent neutrons lost&
Multi- plication factor\\
\hline
100 & 1  & 50 & 20 & 53.9 & 7.0  & 20.0 & 19.1 & 1.03  \\ 
100 & 1  & 50 & 40 & 57.6 & 5.7  & 19.0 & 17.7 & 1.09  \\ 
150 & 1  & 50 & 20 & 58.7 & 6.9  & 18.2 & 16.2 & 1.11  \\ 
150 & 1  & 50 & 40 & 61.3 & 6.0  & 17.5 & 15.3 & 1.15  \\ 
\hline
100 & 2  & 50 & 20 & 49.1 & 10.3 & 20.2 & 20.4 & 0.94  \\ 
100 & 2  & 50 & 40 & 54.2 &  9.8 & 18.0 & 18.0 & 1.02  \\ 
150 & 2  & 50 & 20 & 54.7 & 10.6 & 18.3 & 16.5 & 1.03  \\ 
150 & 2  & 50 & 40 & 57.8 &  9.8 & 17.4 & 15.0 & 1.08  \\ 
\hline
\end{longtable}
\captionof{table}{Neutron balance for a reactor with beryllium reflector  and molybdenum inner shell}
\end{center}

From Table 7 we conclude that a reactor with 1.0 $m$ cavity radius, 50 $cm$ reflector shell thickness, 1 $cm$ molybdenum inner shell can function with fuel density of 20 $kg/m^3$.  A reactor with 1.0 $m$ cavity radius, 50 $cm$ reflector shell thickness, 2 $cm$ molybdenum inner shell can function with fuel density of 40 $kg/m^3$.

\section{Density and chemical equilibrium calculations}

The primary goal of these calculations is determining the plasma conductivity at different temperatures and pressures.  In order for the magnetohydrodynamic generator to work, plasma conductivity must be sufficiently high.

\subsection{Fuel density inside the reactor}

\hskip.8cm We would like to find out the pressure and molecular composition of the fuel with mass composition of 62\% UF$_4$, 30\% KF and 8\% K inside the reactor.  We are given the fuel density of
$20\ kg/m^3$ to $40\ kg/m^3$ and its temperature of $3,500\ ^o$K.  We use the Rocket Propulsion Analysis (RPA) program \cite{RPA} to solve this problem.  Here, Rocket Propulsion Analysis is not used to calculate anything about the propulsion system itself.  Instead, RPA is used to calculate the chemical equilibrium in the reactor chamber depending on composition and pressure.

We have calculated the pressures for reactor fuel at $3,500\ ^o$K and different densities.  For $20\ kg/m^3$, the pressure is $52.5\ atm$, for $30\ kg/m^3$, the pressure is $78\ atm$, and for $40\ kg/m^3$, the pressure is $104\ atm$.  The size of the pressure vessel required to withstand these pressures would depend on the temperature at which it would be kept during the work of reactor.

\subsection{Plasma expansion}

\hskip.8cm As the plasma leaves the reactor, it expands in a nozzle, with some of the thermal energy being converted into mechanical energy.  Then it enters the magnetohydrodynamic (MHD) generator where its kinetic energy is converted into electrical energy.  Along the way and inside the MHD generator, the plasma is subject to the neutron irradiation from the main reactor.  The main purpose of irradiating the expanding plasma is to keep its temperature high enough for conducting electricity.  Additional benefit of keeping the plasma at high temperature is the increased kinetic energy it will acquire during expansion.

The extra ionization of gas provided by fission fragments also increases conductivity \cite{Conductor}, but in our case the fluxes up to $10^{16}\ n/\left(cm^2 \cdot s\right)$ affect conductivity very slightly \cite[p.120]{Conductor}.  Only the highest flux reactors can achieve neutron fluxes above $10^{15}\ n/\left(cm^2 \cdot s\right)$, \cite[p.3]{Flux}, thus all of ionization would have to come from thermal equilibrium.  The main reason why electrons released by fission product ionization do not contribute to conductivity is the fact that they are absorbed by UF$_4$, UF$_5$ and UF$_6$ as soon as they become free.  The molecules of UF$_4$, UF$_5$ and UF$_6$ have electron affinities of $1.27\ eV$, $3.52\ eV$ and $5.63\ eV$ respectively.

The conductivity of plasma is given by the formula \cite[p.136]{UF4-02}
\be
\label{5.01}
\begin{split}
\sigma&=1.44 \cdot 10^7 \left( \Omega \cdot m \right)^{-1}
\left( \frac{Q_e}{\text{\AA}^2} \right)^{-1}
\left(\frac{T}{1,000\ ^oK}\right)^{-0.5}
\frac{n_e}{n_g}\\
&=
1.44 \cdot 10^7 \left( \Omega \cdot m \right)^{-1}
\left( \frac{Q_e}{\text{\AA}^2} \right)^{-1}
T_3^{-0.5} x_e,
\end{split}
\ee
where
$\sigma$ is the conductivity,
$Q_e$ is the mean collision cross-section between electrons and atoms or ions,
$T_3$ is the temperature in $1,000\ ^o$K,
and $x_e=n_e/n_g$ is the molar fraction of free electrons.
According to Klein,  \cite[p.136]{UF4-02}, $Q_e \thickapprox 10 \text{\AA}^2$.  The exact value of $Q_e$ remains to be determined by more theoretical and possibly experimental research.

 At $2,400-2,900\ ^o$K, the conductivity can be approximated by
\be
\label{5.02}
\sigma=.85 \left( \Omega \cdot m \right)^{-1}  f_e,
\ee
where $f_e$ is the molar fraction of free electrons in parts per million.

Below we tabulate free electron abundance for two fuel compositions at different temperatures and pressures of 2 $atm$ and 0.4 $atm$. Composition A contains 62\% UF$_4$, 30\% KF and 8\% K by mass. Composition B 62\% UF$_4$, 38\% KF by mass. The results have been obtained by a Rocket Propulsion Analysis calculation \cite{RPA}.\\
\begin{tabular}{|p{0.9cm}|p{1.3cm}|p{1.3cm}|p{1.3cm}|p{1.3cm}|}
  \hline
   T, $^o$K &
   \multicolumn{4}{|p{5.2 cm}|}{Free electron abundance, ppm for composition and pressure}\\
   & \multicolumn{2}{|p{2.6 cm}|}{Composition A}
   & \multicolumn{2}{|p{2.6 cm}|}{Composition B}\\
   &2 $atm$ & 0.4 $atm$& 2 $atm$ &0.4 $atm$\\
  \hline
  1,800 & 0   & 0.1 & 0   & 0 \\
  1,900 & 0   & 0.5 & 0   & 0 \\
  2,000 & 0.3 & 1.5 & 0   & 0 \\
  2,100 & 0.9 & 4.2 & 0   & 0 \\
  2,200 & 2.3 & 11  & 0   & 0.2 \\
  2,300 & 5.5 & 27  & 0.1 & 0.5 \\
  2,400 & 12  & 58  & 0.3 & 1.4 \\
  2,500 & 26  & 119 & 0.6 & 3.5 \\
  \hline
\end{tabular}
\begin{tabular}{|p{0.9cm}|p{1.3cm}|p{1.3cm}|p{1.3cm}|p{1.3cm}|}
  \hline
   T, $^o$K &
   \multicolumn{4}{|p{5.2 cm}|}{Free electron abundance, ppm for composition and pressure}\\
   & \multicolumn{2}{|p{2.6 cm}|}{Composition A}
   & \multicolumn{2}{|p{2.6 cm}|}{Composition B}\\
   &2 $atm$ & 0.4 $atm$& 2 $atm$ &0.4 $atm$\\
  \hline
  2,600 & 51  & 225 & 1.4 & 8.7 \\
  2,700 & 95  & 395 & 3.2 & 21 \\
  2,800 & 166 & 645 & 6.9 & 50 \\
  2,900 & 277 & 991 & 15 &  111 \\
  3,000 & 437 & 1,440 & 30 & 233 \\
  3,100 & 658 & 2,020 & 61 & 453 \\
  3,200 & 947 & 2,740 &120 & 814 \\
  3,300 & 1,310 & 3,630 &223 & 1,370 \\
  \hline
\end{tabular}
  \captionof{table}{Free neutron content of gaseous fuel at different temperatures}
\vskip.5cm
As it can be seen from the above table the composition without free potassium performs much worse than the composition containing free potassium.  In order for the MHD generator to work, we need the gas entering it to have resistivity of $0.016\ Ohm \cdot m$ at most \cite[p.4]{UF4-01}.  Thus, using formula (\ref{5.02}), one can evaluate that the free electron concentration must be at least $50\ ppm$.

\hskip.8cm For pressures of $2\ atm$ and higher, the expanding fuel can be heated by a neutron flux.  The region where expansion accompanied by nuclear heating takes place is called the \textbf{inner expander}.  From the data presented in Table 8, it follows that the gaseous fuel at $2\ atm$ pressure must have a temperature of $2,600\ ^o$K.  According to a Rocket Propulsion Analysis calculation \cite{RPA}, if the fuel expands adiabatically from $78\ atm$ pressure
(corresponding to $30\ kg/m^3$ density) to $2\ atm$ pressure,
then its final temperature would be $2,090\ ^o$K.
Thus, heating by neutron irradiation and fission must provide enough energy to increase the temperature by extra $510\ ^o$K.

\subsection{Additional expansion and heating of gaseous fuel}

\hskip.8cm The gaseous fuel leaves the \textbf{inner expander} with velocity $1.2\ km/s$, internal pressure $2\ atm$, temperature $2,600\ ^o$K, and density $1.02\ kg/m^3$.  Previous models
discussed in the literature \cite{MHD03,UF4-01,UF4-02}
suggest reducing internal gas pressure to $0.65\ atm$. In our model, the pressure is reduced to $0.4\ atm$.  As we see from Table 8, in order for the gaseous fuel to have electron density of at least $50\ ppm$, it has to have a temperature of $2,400\ ^o$K at $0.4\ atm$ internal pressure.

\hskip.8cm According to a Rocket Propulsion Analysis calculation \cite{RPA} if gaseous fuel adiabatically expands from $2\ atm$ to $0.4\ atm$, its temperature will decrease from $2,600\ ^o$K to $2,050\ ^o$K.  Thus, in order for MHD generator to work, the gas would have to be heated by at least $350\ ^o$K.  Neutron and fission heating at densities under 1 $kg/m^3$ is impossible, and the use of generated electricity for heating is wasteful.  One option, which should be explored, is \textbf{mechanical heating}.  Mechanical heating uses some of the the kinetic energy of the gas to heat it.  It is achieved by placing obstacles like tungsten or tantalum carbide rods in the path of the gas stream.  These obstacles will cause some of the gas kinetic energy to be transferred into heat.

\hskip.8cm The MHD generator will also convert some of the gaseous fuel's kinetic energy into thermal energy by means of resistive heating of the gas.  The gaseous fuel will enter the MHD generator as a turbulent flow with velocity $1.26\ km/s$, internal pressure $0.4\ atm$, temperature $2,400\ ^o$K, and density $0.22\ kg/m^3$.  The MHD generator will extract about 50\% of the gas kinetic energy and convert it into electricity \cite[p.102]{MHD03}.  The gas will also expand to internal pressure of $0.08\ atm$ in the generator itself \cite[p.8]{UF4-01}.  The gas will slow down and retain 15\% of its kinetic energy \cite[p.8]{UF4-01}.  Since 50\% of the gaseous fuel's kinetic energy will turn into electrical energy and 15\% of kinetic energy will remain, 35\% of initial kinetic energy will become thermal energy of the gas.  The thermal energy supplied to the gaseous fuel in MHD generator will thus be $.35 \cdot v^2/2$ which is $280\ J/g$.  According to a Rocket Propulsion Analysis calculation, the fuel gas will leave the MHD generator at $2,920\ ^o$K.  Some of the waste heat may be used to heat liquid UF$_4$ as it begins the next cycle.

\subsection{Gaseous fuel condensation}
\hskip.8cm The gaseous fuel containing 62\% UF$_4$, 30\% KF and 8\% K by mass enters condensing radiator at $0.1\ atm$ pressure.  As the mixture cools it begins to condense.  As it condenses, there will be equilibrium between UF$_3$, UF$_4$, K and KF
given by the following diagram:
\be
\label{5.03}
K+ U F_4 \leftrightharpoons KF+ UF_3.
\ee
According to Rocket Propulsion Analysis calculation, as the temperature falls to $1,490\ ^o$K, 34.5\% of the fuel condenses as UF$_3$ crystal.
As the temperature drops to $1,300\ ^o$K, less than a quarter of potassium remains unbound to F.
The exact phase diagram of solution containing UF$_4$, UF$_3$, KF, and K remains to be \emph{determined experimentally}.

\hskip.8cm We would suggest dissolving the fuel in liquid MgF$_2$ and/or $^{88}$SrF$_2$.  Most of the solvent would remain liquid throughout the cycle and never enter the reactor cavity.  Finding the best solvent and accounting for all fission product effects remains \emph{an open problem.}

\subsection{Electrical efficiency estimation}

\hskip.8cm
Now we give approximate estimate on the thermal energy which is supplied to fuel in each cycle and on the amount of thermal energy which is converted to
electric energy by the MHD generator.  As the fuel begins the cycle, the 62\% UF$_4$ fraction and the 30\% KF fraction enter the reactor as liquids.  The heat of vaporization of KF is $4.2\ kJ/g$ and the heat of vaporization of UF$_4$ is $1.0\ kJ/g$ \cite{RPA}.  Thus for the whole fuel mass, vaporization takes 1.9 $kJ/g$.  According to a Rocket Propulsion Analysis calculation, heating the gaseous fuel to $3,500\ ^o$K takes another $1.0\ kJ/g$.  As the fuel expands, it is heated by fissions caused by neutron flux raising the final temperature from  $2,090\ ^o$K to $2,600\ ^o$K, which takes another $0.3\ kJ/g$.  The total amount of thermal energy supplied to the fuel during each cycle is thus about
$1.9+1.0+.3=3.2\ kJ/g$.

\hskip.8cm According to a Rocket Propulsion Analysis calculation as the fuel expands from $78\ atm$ pressure corresponding to $30\ kg/m^3$ density to $2\ atm$ pressure, it obtains $0.75\ kJ/g$ in kinetic energy.
Additional expansion from $2\ atm$ pressure to $0.4\ atm$ pressure will bring only
$0.05\ kJ/g$ in kinetic energy, since most of the energy of expansion will be returned as thermal energy by mechanical heating.
Previous studies \cite[p.102]{MHD03} mention isentropic efficiency of 50\% for a Space based MHD generator.  Thus, $0.8\ kJ/g$ kinetic energy will yield $0.4\ kJ/g$ electrical energy.  The overall efficiency of the power plant consuming $3.2\ kJ/g$ thermal energy per cycle and producing $0.4\ kJ/g$ electrical energy per cycle is 12.5\%.

\subsection{Radiator: the heat sink}
\hskip.8cm Any heat engine needs a heat sink.  A heat engine with electrical power output $P_e$ and efficiency $\eta$ needs a heat sink with thermal power
\be
\label{5.04}
P_s=\frac{1-\eta}{\eta} P_e.
\ee
For a 12.5\% ($\eta=0.125$) efficient system, the heat sink has to reject almost 7 times more power than the electric generator produces.  In our case, the heat sink is a \textbf{radiator}.  A radiator is a structure composed of metal tubes linked by metal bends.  The working fluid passes through the tubes and radiates its thermal energy.  In our case, the working fluid composed of UF$_4$, KF, and K changes phase from gas to liquid within the radiator.  The structure also has many faucets which are robotically controlled -- if one or more tubes leak, than the working fluid flow is shut from them.

\hskip.8cm The main parameter for radiator performance is the physical weight of the system needed to reject a unit of thermal power measured in $kg/kW$.  Due to the Stefan-Boltzmann Law, ``the absolute measure of the heat rejection effectiveness is the $T^4$ dependence" \cite[p.7]{UF4-01}.  Since UF$_4$ has much higher condensation point then other working fluids, it offers an ``order-of magnitude
improvement in radiator performance" \cite[p.7]{UF4-01}.  The plot showing radiator performance is given in Fig. 6 below \cite[p.17]{UF4-01}:
  \begin{center}
  \includegraphics[width=12cm]{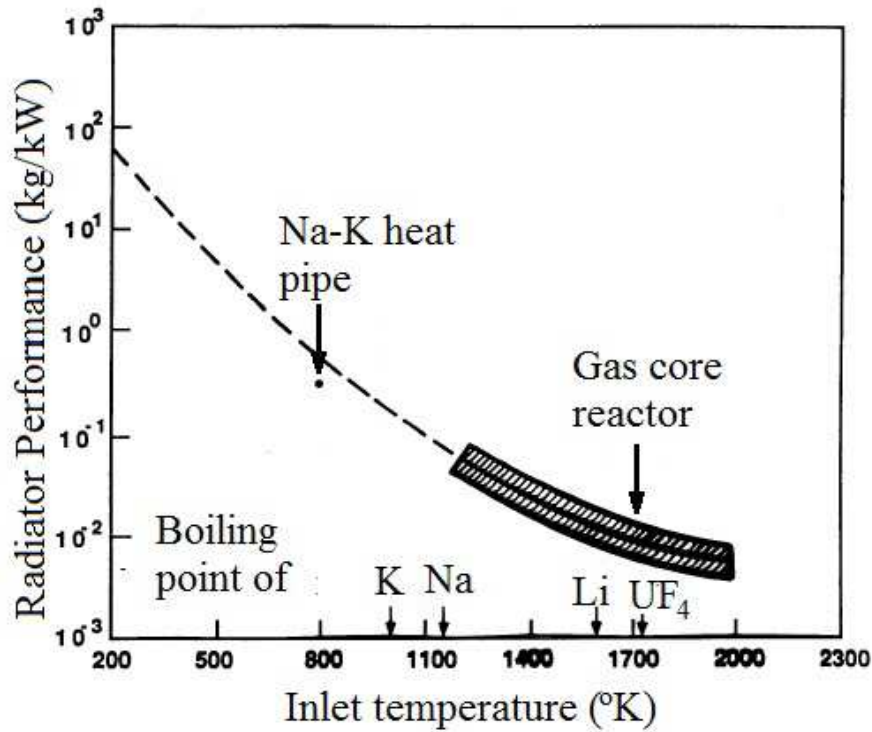}
  \vskip0cm
  \captionof{figure}{Radiator performance}
  \end{center}

\section{Economic Considerations}
\subsection{State of the art deep space launch costs}
\hspace{.8cm} Overall, from 1957 to 2015, there have been 5,510 space launches in the world of which 5,046 were successful.  The decade with most (1,231) space launches was 1970s  \cite{SLR}.  During the years 1990 -- 2010, 16,200 tons of payload have been launched into space --mostly Low Earth Orbit (LEO) and 65\% of that mass returned to Earth with Space Shuttle \cite{TotM}.

\hspace{.8cm}Saturn V Rocket used in Moon expedition costs \$110 million in 1970 dollars \cite[p.406]{Saturn}, which is \$680 million in 2015 dollars.  It could have placed 108 tons into LEO \cite[p.43]{Saturn1}.  Thus, by 1969, very heavy launch vehicles could place payload into orbit at \$6,300 per kg in 2015 dollars.  Most experts in early 1970s believed that in a few years, the cost of placing payload into orbit would be \$2,200 per kg in 2015 dollars \cite[p.35]{Shuttle}.  By 1980s space launch should have cost \$400 -- \$600 per kg payload in 2015 dollars \cite[p.viii]{guide}.  Unfortunately, the launch costs have not improved since 1969.  In 2002, the cost of LEO delivery has been \$6,900 per kg for Delta4 Heavy and \$18,400 per kg for Titan 4B \cite[p. 211]{Cost1}.  In 2016, Ariane 5 ECA delivers payload to LEO for \$8,500 per kg \cite[p.74]{Raport01}, while Falcon Heavy delivers payload to LEO for \$5,100 per kg \cite[p.94]{Raport01}.  The combination of bright predictions in early 1970s and quite slow progress by 2016 makes the time frame for future development uncertain.

\subsection{GCR cost estimation}
\hskip.8cm We would like to estimate the cost of a Liquid-Vapor Core Reactors with MHG described in the previous section.  The concept reactor  \cite{MHD03} has a mass of $200\ tons$ and electrical power of $200\ MW$.  Given that our system has 12.5\% efficiency, the thermal power would have to be $1.6\ GW$.
We would like to estimate material inventory needed for the reactor.
\begin{enumerate}
  \item Uranium $^{\text{235}}$U:\\
  The reactor should work for $5$ years at the thermal power of $1.6\ GW$, producing
  $2.5 \cdot 10^{17}\ J$ of thermal energy.  Each $^{\text{235}}$U fission produces $190\ MeV$ \cite[p.10]{neutronicsA}, thus $^{\text{235}}$U produces
  \be
  \label{6.01}
  1.9 \cdot 10^8 \frac{eV}{\text{atom}} \cdot \frac{1.602 \cdot 10^{-19} J}{1\ eV}
  \cdot \frac{6.022 \cdot 10^{23} \text{atom}}{ \text{mole}} \cdot
  \frac{1 \text{mole} }{235\ g}=78 \cdot 10^9\ \frac{J}{g}=78 \ \frac{GJ}{g}.
  \ee
  The total amount of $^{\text{235}}$U consumed by the reactor would be
  \be
  \label{6.02}
  \frac{2.5 \cdot 10^{17}\ J}{78 \cdot 10^9\ J/g}=3.2 \cdot 10^6\ g= 3.2 \ ton.
  \ee
  As we have mentioned earlier, as much as 60\% of $^{\text{235}}$U would be discarded with the growth of $^{\text{236}}$U and fission product contamination.  Thus, the reactor would need $8.0\ tons$ of $^{\text{235}}$U.  If the uranium used in the reactor is 99\% enriched, and the fuel consists of 62\% UF$_4$ and 30\% KF and 8\% K, then $^{\text{235}}$U makes up 46\% of fuel by mass.  Thus the total mass of the fuel is $17\ tons$.
  In our system, one gram of input $^{\text{235}}$U would produce $1,080\ KWh$ electricity.
  \item Beryllium:\\
  The reactor gas core is a cylinder surrounded by a 2 $cm$ thick molybdenum shell.  The shell is surrounded by a 50 $cm$ thick liquid beryllium shell.  The spherical cavity inside neutron reflector shell would have radius $r_0=1.2\ m$.  The outer radius of the shell is $r_1=1.72\ m$.  The total volume of beryllium oxide shell denoted by $V_{Be}$ is given by
  \be
  \label{6.03}
  V_{Be}=\frac{4 \pi}{3} (r_1^3-r_0^3)=13.7 m^3.
  \ee
  The density of liquid beryllium is $1.69\ g/cm^3$, thus the mass of beryllium shell is $23.2\ tons$.
  \item Tungsten:\\
  Most of the rest of the engine as well as the heat radiation system is made from tungsten -- thus total tungsten mass is about $140\ tons$.
\end{enumerate}

\hspace{.8cm} Now we can look at the prices of materials used.  Beryllium costs up to \$500/kg  \cite{Beryllium}.  Tungsten costs up to \$40/kg \cite{tungsten}.  The price of $^{\text{235}}$U has two components -- the price of natural uranium and the price of isotope separation.  The separation work is measured in \textbf{separative work units} (SWUs) \cite[p.2]{SWUprice}.  In order to separate the feed mass $M_f$ of uranium with $^{\text{235}}$U fraction $x_f$ into
product with mass $M_p$ of uranium with $^{\text{235}}$U fraction $x_p$ and
waste with mass $M_w$ of uranium with $^{\text{235}}$U fraction $x_w$, we need to apply the separative work given by \cite[p.2]{SWUprice}:
\be
\label{6.04}
\text{Separative Work}=M_p V\left(x_p\right)+M_w V\left(x_w\right)-
M_f V\left(x_f\right),
\ee
where
\be
\label{6.05}
V\left(x\right)=(1-2x) \ln \left(\frac{1-x}{x} \right).
\ee
Thus, in order to separate $M_f=200\ kg$ of natural uranium containing
$x_f=7.1 \cdot 10^{-3} \ ^{\text{235}}$U
into
$M_w \sim 199\ kg$
of depleted uranium containing
$x_w=2.1 \cdot 10^{-3} \ ^{\text{235}}$U and $M_p=1.01\ kg$ enriched uranium containing
$x_p=0.99\ ^{\text{235}}$U, we need $262\ kg$ separation work, which is 262 SWU's.
In 2014, one SWU costed \$140 \cite[p.2]{SWUprice}.  At present, natural uranium
costs \$112 per $kg$ \cite[p.2]{SWUprice}, but the prices are subject to fluctuation.  In 2007, uranium prices reached \$300 per $kg$ \cite{U2007}.  Uranium can be extracted from sea water at \$260 per $kg$, \cite{USee} which should stabilize the price.  In order to obtain $1\ kg$ of $^{\text{235}}$U, we need $200\ kg$ of natural uranium at \$260 per $kg$ and 262 SWU at \$140 each -- \$89,000 total.  The total material cost is \$850 million.

\hspace{.8cm}The capital costs of nuclear power systems in the year 2000 was \$2.4 per $Watt$ in 2015 dollars.  The price in 2015 was \$4.6 per $Watt$ \cite[p.98]{NCost} due to new safety regulations.  Given that in the deep space, safety is less important we can assume the system cost of \$2.4 per $Watt$ which is \$480 million total.

\hspace{.8cm}The reactor can be launched into deep space by Falcon 9 rocket, or its future derivative.  Given that nuclear fuel must be enclosed in special capsules, the total payload should be about 250 tons.  The cost of launching 250 tons of payload into deep space by modern Falcon 9 rocket is \$3.2 billion \cite[p.19]{Raport01}.

\hspace{.8cm}The grand cost of the 200 $Megawatt$ system is \$850 million plus \$240 million plus \$3.2 billion, which is \$4.2 billion total or \$21 per $Watt$.  Solar power systems which cost at least \$1,000 per $Watt$ can not compete with nuclear power.

\subsection{The best use of uranium resources}

\hskip.8cm During the 20$^{\text{th}}$ century, nuclear technology has played an important role in development of Humankind.  The existence of nuclear weapons was a major factor during Cold War and in prevention of major wars after WWII.

\hskip.8cm During the early years of nuclear energy use, it was expected that nuclear power would provide  inexpensive electricity, which did not become the reality.  Nuclear power production has experienced exponential growth between 1959 and 1989, then slow growth in 1990s and has stabilized in 2000s \cite[p.109]{Uranium}.  In USA, the total nuclear power stabilized at about $100\ GW$ by the end of 1980s and has not grown up to 2017.  According to the leading experts in 1972, the total installed nuclear power should have been $508\ GW$ by 1990 and $1,200\ GW$ by 2000 \cite[p. 14]{allnuclear}.

\hskip.8cm In thermal reactor based electric power stations, which use almost exclusively the rare isotope $^{235}$U, one kilogram of the natural uranium produces about 40,000 $kWh$ electricity \cite[p.77]{Uranium}.  Fast breeder reactors which could utilize all of the uranium are extremely expensive \cite[p.179]{Kotov}.  Reasonably assured resources of uranium consist of 4.6 million tons of uranium \cite[p.21]{Uranium} from which 27,000 tons of $^{235}$U can be separated.  In terms of global energy reserves, $^{235}$U contains 21 times less energy than coal \cite[p.17]{WorldEnergy}.  Thus, nuclear energy hardly has future perspectives on the Earth.

\hskip.8cm Nevertheless, fission energy can be a key factor in the initial stages of Solar System colonization.  As we have shown in this work, nuclear energy is much less expensive than solar energy in deep space.  Moreover, it can be used far from the Sun, where solar power is very weak.  Delivery of chemical fuel into Deep Space is prohibitively expensive, yet $^{235}$U has thermal energy potential of 68 $GJ/g$, thus it can store compact energy for deep space missions.  Finally, even though fast breeder and thorium based reactors have been extremely expensive on the Earth due to safety concerns, they may be inexpensive in deep space.

\section{Summary}

\hskip.8cm Several options have been considered in the past for deep space propulsion.   Even though chemical fuel rockets are very well developed, the cost of bringing enough chemical fuel into deep space is prohibitive.  The best option for deep space propulsion would be magnetoplasmodynamic thrusters which use electric energy to eject lithium plasma at 15 to 60 $km/s$.  These thrusters have power of $500\ kW$ each and electrical efficiency of 40\% to 60\%.  The only realistic option for supplying these thrusters with electric power would be nuclear reactors and nuclear fuel from the Earth.  The nuclear reactors and spaceships used in deep space must be as compact as possible,  since it costs at least \$20,000 per kilogram to deliver any payload into deep space.

\hskip.8cm The power requirement for magnetoplasmodynamic thrusters is at least $45\ W$ per kilogram of loaded spaceship mass.  Thus, nuclear electric power stations with maximal possible \textbf{specific power} (power per unit mass) are needed.  The system we describe in this work is the Gas Core Reactor (GCR) with magnetohydrodynamic (MHD) generator.  One such system proposed to power a heavy space transport has electrical power output of $200\ MW$ and a mass of $200\ tons.$  It would operate 5 years using $9.3\ tons$ of $^{235}$U, which costs \$89 per gram.  The overall cost of such station is estimated at \$4.2 Billion.

\hskip.8cm The Gas Core Reactor (GCR) is a reactor in which the fuel generates thermal energy in gaseous state.  In our concept, the fuel consists of a mixture of $^{235}$UF$_4$, KF, and K.  The mixture enters the reaction chamber in gaseous state at $40\ atm$ to $60\ atm$ pressure.  The reaction chamber is surrounded by a shell of liquid beryllium neutron reflector in a molybdenum – 92 container.  Within the reaction chamber, the working fluid is heated by fission to 3,500$^o$K.  Then the working fluid  expands through a nozzle and enters a MHG.  Finally, the working fluid is cooled and liquefied in a condensing radiator.  Then it enters the reactor, where it is vaporized at a high pressure.  Then it returns to reaction chamber and the cycle repeats itself.

\hskip.8cm We have performed a numerical simulation describing neutronic performance of the gas core reactor.  Each simulation consisted of tracing the simulated paths of 5,000 to 20,000 neutrons within the reactor.  From these paths, criticality as well as the rates of neutron absorption and fission for different reactor regions have been estimated.  The simulation error is proportional to the inverse of the square root of the number of neutrons: Err$\sim 1/\sqrt{N_n}$.

\hskip.8cm Each neutron path is simulated from emission to absorption or escape.  A neutron is emitted from a random point within the reactor cavity.  It is emitted in a random direction. It has a random energy chosen from an energy distribution (see formula (\ref{3.03})).  As the neutron travels within the reactor, it undergoes \textbf{events}.  The following constitutes an \textbf{event}: 1) neutron moving from one region of reactor to another, 2) neutron scattering, 3) neutron absorbtion, 4) neutron escape from the reactor. The last two events end the neutron path.  The reactor is divided into three regions.  The first and innermost region is a sphere of gaseous fuel.  The second region is a thin shell of refractory material region. The third region is a thick shell of Beryllium reflector.

\hskip.8cm The probabilities of absorbtion or scattering taking place within a region of reactor are determined by neutron absorbtion and scattering cross -- sections.  These cross -- sections are energy dependent.  Both inelastic and elastic scattering events are considered in the simulation.  A typical neutron undergoes several hundred collisions within the Beryllium reflector region before being absorbed or escaping the reactor.  About 90\% of neutrons thermalise before escaping or being absorbed.

\hskip.8cm After simulating the paths of the neutrons, we calculate the reactor criticality.  The \textbf{regeneration factor} $\eta$ is the quotient of the number of neutrons produced from fission to the number of neutrons absorbed in fuel.  In the gaseous fuel region, $\eta=1.96$, thus by multiplying the fraction of the neutrons which end up absorbed in the fuel region by $1.96$, we obtain criticality.  The criticality has to be corrected for two effects which almost cancel each other out: neutron absorbtion by fission products and neutron multiplication in Beryllium.

\hskip.8cm The first simulation deals with performance of Beryllium neutron reflector.  The fractions of neutrons which leak from reactor for different values of reactor cavity radius and reflector thickness are presented in Table 4.  Reflector thickness of $50\ cm$ produces leakage of about 0.9\%.  The values of reflectance for an infinite Beryllium medium surrounding a cavity of a given radius are presented in Table 5.  These reflectances are 86\% to 90\%.  The second simulation shows that tungsten can not be used as refractory material, since it's neutron absorbtion is too high.  The results are in Table 6.  The third simulation shows that $^{92}$Mo is a good choice for refractory material.

\hskip.8cm Next, we calculate the density and resistivity of the gaseous working fluid entering the MHD generator.  We show that in order for the plasma to be conductive it has to be kept at a temperature of at least 2,600 $^o$K.  We estimate the efficiency of the power station at 12.5\%.  Then we describe the radiation heat sink and show that it has to be kept at high temperature of about 1,500$^o$K.  A colder heat sink would enable the station to generate electricity with greater efficiency, but it would take too much mass.

\end{document}